\documentclass[runningheads]{llncs}
\usepackage{amsmath}
\usepackage{amssymb}
\usepackage{bm}
\usepackage{mathtools}
\usepackage{cite}
\usepackage{subcaption}
\usepackage{graphicx}
\usepackage{float}
\usepackage{booktabs}
\usepackage{multirow}
\usepackage[table]{xcolor}
\usepackage{siunitx}
\usepackage[ruled]{algorithm2e}

\usepackage[tableposition=top, labelfont=bf, labelsep=period]{caption}
\usepackage{tikz}
\usetikzlibrary{positioning, arrows}

\newcommand{\x}{\ensuremath{\bm{x}}}
\newcommand{\X}{\ensuremath{\bm{X}}}
\newcommand{\z}{\ensuremath{\bm{z}}}
\newcommand{\xhr}{\ensuremath{\x_{\text{\tiny HR}}}}
\newcommand{\xlr}{\ensuremath{\x_{\text{\tiny LR}}}}

\newcommand{\eqendp}{\,\text{.}} 
\newcommand{\eqendc}{\,\text{,}} 

\definecolor{cb-green-sea}  {RGB}{  0, 146, 146}
\definecolor{cb-burgundy}   {RGB}{146,   0,   0}

\begin{document}

\title{Cascaded Latent Diffusion Models for High-Resolution Chest X-ray Synthesis}
\titlerunning{Cascaded Latent Diffusion Models for Chest X-ray Synthesis}
%
\author{Tobias Weber\inst{1,2,3}\orcidID{0000-0002-5430-2595} \and
Michael Ingrisch\inst{2,3}\orcidID{0000-0003-0268-9078} \and
Bernd Bischl\inst{1,3}\orcidID{0000-0001-6002-6980} \and
David Rügamer\inst{1,3}\orcidID{0000-0002-8772-9202}}

\authorrunning{T. Weber et al.}

\institute{Department of Statistics, LMU Munich\\
\and
Department of Radiology, University Hospital, LMU Munich\\
\and
Munich Center for Machine Learning (MCML)\\
\email{tobias.weber@stat.uni-muenchen.de}
}


\maketitle              
%

\begin{abstract}
While recent advances in large-scale foundational models show promising results, their application to the medical domain has not yet been explored in detail.
In this paper, we progress into the realms of large-scale modeling in medical synthesis by proposing \textit{Cheff} - a foundational cascaded latent diffusion model, which generates highly-realistic chest radiographs providing state-of-the-art quality on a 1-megapixel scale.
We further propose \textit{MaCheX}, which is a unified interface for public chest datasets and forms the largest open collection of chest X-rays up to date.
With \textit{Cheff} conditioned on radiological reports, we further guide the synthesis process over text prompts and unveil the research area of report-to-chest-X-ray generation.

\end{abstract}
%
%

%
\section{Introduction}

Chest X-ray examinations are one of the most common, if not the most common, procedures in everyday clinical practice.
This not only enables the collection of large databases but paves the way for lots of great opportunities to advance medical AI assistance in clinical workflows.
Automated lung disease diagnosis \cite{rajpurkar_chexnet_2017}, for example, can help to accelerate an examination and assist radiologists by reducing human error in a high-stress environment.
Despite the availability of public datasets, many challenges in practical usefulness remain, partly induced due to inherent class imbalances and noisy labels.
One possible approach to tackle this problem is by synthesizing underrepresented classes via generative modeling \cite{sundaram_gan-based_2021}.
Generative models provide various other opportunities to improve clinical routines by, e.g., modeling the characteristics of diseases in different degrees using the patient's original thoracic scan \cite{weber_implicit_2022} or suppressing bones to enhance soft tissue within the lung \cite{han_gan-based_2022}.

The basis for these methods is a stable and high-quality synthesis process. While previous evaluations in clinical practice focus on generative adversarial networks (GANs; see, e.g., \cite{segal_evaluating_2021}),
the trend of generative architectures moves away from an adversarial setting. Current state-of-the-art approaches tend to employ large-scale autoencoders with a focus on generating a prior in their latent space \cite{ramesh_zero-shot_2021, esser_taming_2021, rombach_high-resolution_2022}.
Moreover, diffusion models \cite{sohl-dickstein_deep_2015, ho_denoising_2020} have shown immense potential in image synthesization \cite{dhariwal_diffusion_2021, nichol_improved_2021} resulting in a variety of proposed large-scale models including GLIDE \cite{nichol_glide_2022}, DALL·E 2 \cite{ramesh_hierarchical_2022}, Imagen \cite{saharia_photorealistic_2022} or Stable Diffusion \cite{rombach_high-resolution_2022}.

Recently, GoogleAI released an API \textit{CXR Foundation Tool} offering chest radiograph embeddings based on a contrastive neural network \cite{sellergren_simplified_2022}.
The underlying training dataset contains a significant proportion of non-disclosed proprietary data, though.
The imminent risk of biases in foundational chest X-ray models has been recently discussed in \cite{glocker_risk_2022} warning about the consequences in real-world applications.
While these biases are inherent in the existing data, closed-source analyses make it difficult for the research community to counteract these problems. 
\begin{figure}[t]
\centering
    \includegraphics[width=\textwidth]{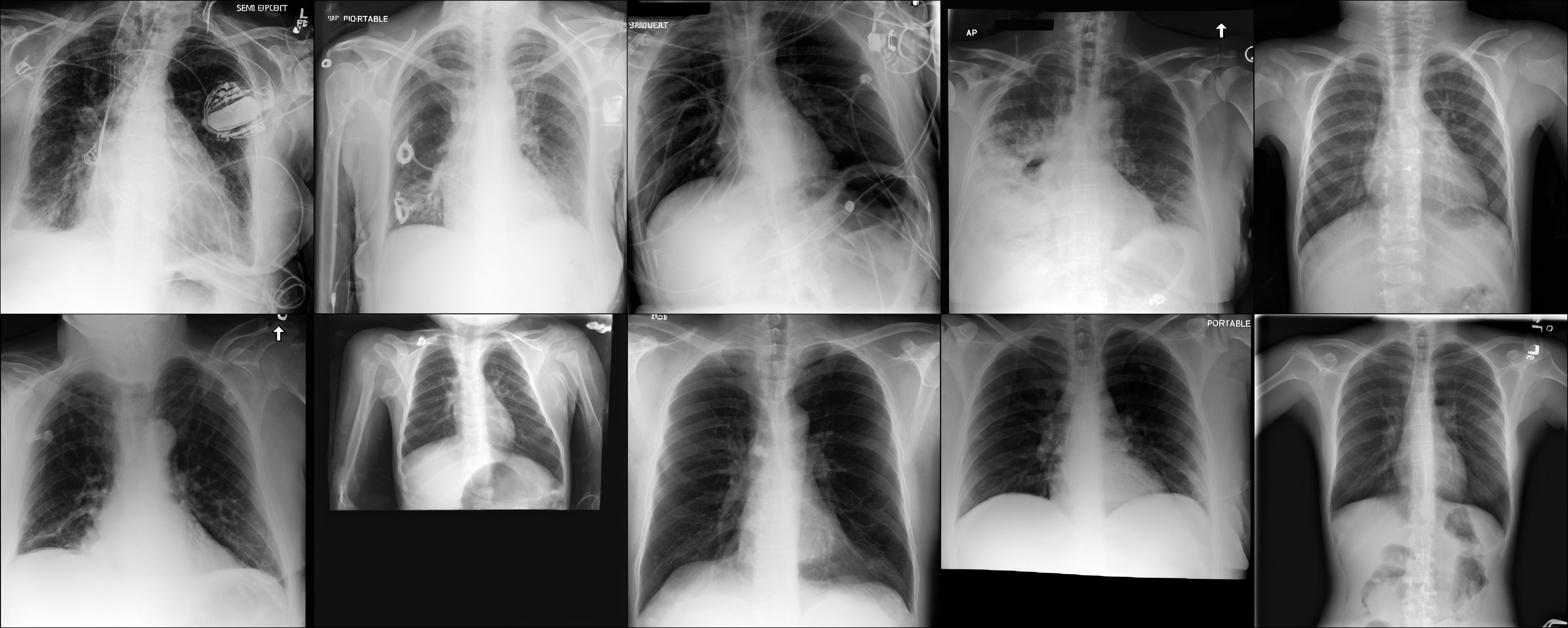}
    \caption{High-resolution synthetic chest X-rays generated with our cascaded latent diffusion pipeline \textit{Cheff}.}
    \label{fig:diff-introgrid}
\end{figure}
\paragraph{Our Contributions}
In this paper, we adapt large-scale learning concepts and models of general computer vision synthesis to chest radiography.
To make this possible, our first contribution is to provide a unified interface to a massive collection of public chest X-ray datasets (\textit{MaCheX}) with over 650,000 scans.\setcounter{footnote}{0}\footnote{\textbf{{MaChex:}} https://github.com/saiboxx/machex}
This dataset allows the training of \textit{Cheff}, a cascaded \textbf{che}st X-ray latent di\textbf{ff}usion pipeline, our second contribution.\footnote{\textbf{{Cheff:}} https://github.com/saiboxx/chexray-diffusion} The basis for \textit{Cheff} are two foundational chest X-ray models: (i) an autoencoder for obtaining chest X-ray embeddings and (ii) a super-resolution diffusion model for refining low-resolution scans.
A task-specific diffusion process in the latent space of the autoencoder leads to high-fidelity and diverse synthesis (see, e.g., Figure~\ref{fig:diff-introgrid}).
This further enables variable conditioning mechanisms while reducing computational training costs due to the usage of shared foundational models in the full pipeline.
Finally, our model provides realistic report-to-chest-X-ray conversion by applying our model stack in a text-to-image setting by conditioning on radiologists' reports.
Overall, our approach pushes the limits in state-of-the-art  synthesis of radiological images and fosters development of downstream clinical AI assistance tools by making both our model and dataset publicly accessible.
\section{Related Work} \label{sec:related}

In recent years, the rising popularity of generative models resulted in various approaches focusing on chest X-ray synthesis.
\cite{segal_evaluating_2021} investigate the class-guided synthesis of targeted pathologies over classifier gradients.
Their training is, however, only based on a single-center dataset while using a progressive-growing GAN (\!\cite{karras_progressive_2018}; PGAN) known to be less capable than other more recently proposed algorithms.
\cite{weber_implicit_2022} obtain chest scan embeddings by applying GAN inversion to the generator of \cite{segal_evaluating_2021} and investigate characteristics of pathologies in this latent space.
\cite{han_breaking_2020} utilize privacy-free synthetic chest X-rays for federated learning to ensure data protection across hospitals.
\cite{tang_disentangled_2021} proposes a decomposition of diseases, ultimately producing anomaly saliency maps, by adding a second generative network exclusive for abnormal scans to produce a residual for a healthy version of this scan. 
Other use cases of generative methods include bone suppression \cite{han_gan-based_2022} or improving classifier performance on underrepresented classes \cite{sundaram_gan-based_2021}.

Apart from the classic adversarial literature, there is a surge in diffusion model research that also deals with chest X-rays.
\cite{wolleb_diffusion_2022} propose a routine training on normal data and then utilizes a partial denoising process to detect anomalies.
\cite{chambon_adapting_2022} fine-tune the foundational Stable Diffusion on a few thousand scans.
While the synthesized scans lack a realistic style, it allows for a first glimpse into how report-to-scan generation can be used in medical imaging.
\cite{packhauser_generation_2022} promote the usage of diffusion models over PGANs by fitting a latent diffusion model (LDM; \cite{rombach_high-resolution_2022}) on chest X-rays and observing a performance boost in a synthetically aided classifier.
Additionally, the synthesizing process ensures anonymity with respect to the patient retrieval problem by employing a privacy-enhancing sampling algorithm \cite{packhauser_deep_2022}.
A different strategy involves prompting DALL·E 2 to investigate zero-shot capabilities for medical imaging synthesis, however with limited success \cite{ali_spot_2022}.
Generative modeling for automated radiological reports from chest X-rays is another highly active research area (cf. \cite{liu_clinically_2019, liu_exploring_2021}), whereas the topic of creating images from reports is still an underrepresented, if not an unexplored, subject.


\section{Methods} \label{sec:methods}

In the general framework of diffusion models \cite{sohl-dickstein_deep_2015, ho_denoising_2020}, a Markov chain $(\X_t)_{t=1,\ldots,T}$ with joint distribution defined by the density $q(\x_{1:T}) \coloneqq \prod^T_{t=1} q(\x_t | \x_{t-1})$ is used to model a diffusion process of an uncorrupted image $\x_0 \in \mathcal{X}$. This is done by assuming $\X_t \mid \X_{t-1}\!=\!\x_{t-1} \sim \mathcal{N}(\sqrt{1-\beta_t} \x_{t-1}, \beta_t \bm I)$ with time-varying variance $\beta_t \in (0,1)$, which is chosen by a pre-defined variance schedule. In other words, this autoregressive process incrementally applies Gaussian noise to an input $\x_0$ for a number of timesteps $T$.
The learning task then is to revert this so-called \textit{forward process}, i.e., inferring $\x_{t-1}$ from its corrupted version $\x_t$ for $t = 1, ..., T$. For large $T$ the result $\x_T$ will approximate isotropic Gaussian noise and the learned \textit{reverse process} can be utilized as a powerful iterative generative model to model structured information from noise. 

A simplified training objective (c.f. \cite{rombach_high-resolution_2022, ho_denoising_2020}) can be formulated by minimizing
\begin{equation}
    \mathbb{E}_{t\sim U(1,T),\x_0 \in \mathcal{X}, \bm \epsilon_t \sim \mathcal{N}(\bm 0, \bm 1)} \left[ \Vert \bm \epsilon_t - \bm \epsilon_{\theta}(\x_t, t) \Vert^2_2 \right] \eqendc
\end{equation}
where $\bm \epsilon_\theta$ usually is a time-conditional U-net \cite{ronneberger_u-net_2015} parameterized with $\theta$ to predict the noise $\bm \epsilon_t$ in $\x_t$ as a function of $\x_0$  for uniformly drawn $t$ from \{1,\ldots,T\} using the fact that $\x_t = \sqrt{\bar{\alpha}_t}\x_0 + \sqrt{1-\bar{\alpha}_t}\bm \epsilon_t$ with $\bar{\alpha}_t = \prod_{s=1}^t (1-\beta_s)$.

\paragraph{Cheff}

\begin{figure}[b]
\centering
    \includegraphics[width=0.9\textwidth]{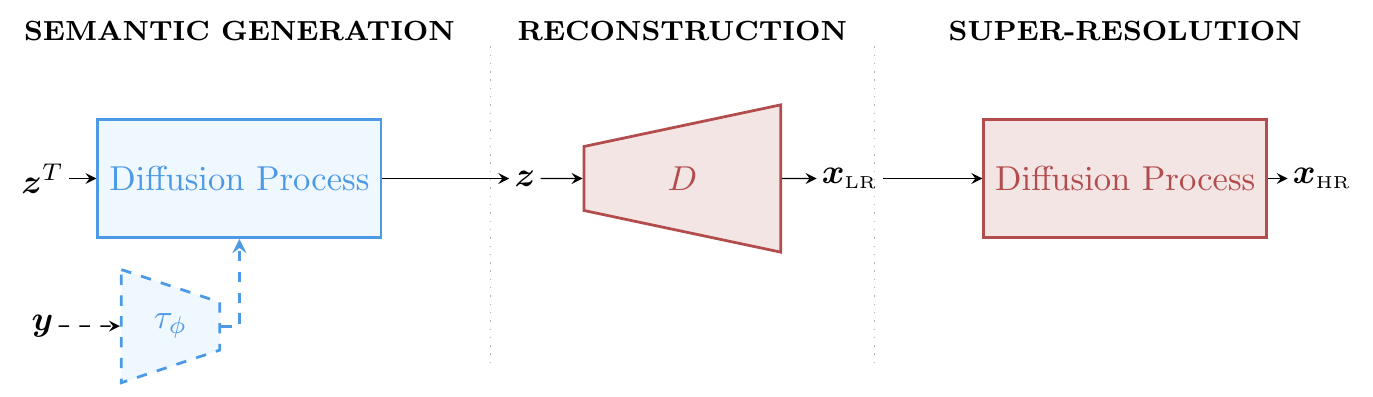}
    \caption{Flow of our cascaded synthesis pipeline. A first diffusion model generates a latent sample $\z \in \mathcal{Z}$ from $\z^T \sim \mathcal{N}(\bm 0, \bm I)$ through its reverse process. Optionally, an embedded conditioning $\tau_\phi(\bm y)$ can be added, where $\bm y$ in our application could, e.g., be pathology labels, radiological reports, or radiologist's annotations. The decoded image $\xlr = D(\z)$ is subsequently refined by a second diffusion model to a high-resolution image \xhr. Foundational models are marked in red, whereas blue components are task-specific.}
    \label{fig:diff-overview}
\end{figure}

We now present our method \textit{Cheff} to iteratively generate high-resolution images. Our multi-stage synthesis approach can be decomposed into three cascading phases:
(i) Modeling a diffusion process in latent space, (ii) translating the latent variables into image space with a decoder, (iii) refinement and upscaling using a super-resolution diffusion process.
Phase (i) and (ii) together define an LDM. Figure~\ref{fig:diff-overview} summarizes this multi-stage approach again.

Every model can be trained in an encapsulated fashion as described in the subsequent subsections.
In the following, we denote \xhr\ as a high-resolution image and \xlr\ as its lower-resolution variant (in our model stack these are 1024 and 256 pixels, respectively).
We start by explaining the second (reconstruction) part of our model first.
\paragraph{Autoencoder training}
It has been shown that applying diffusion in a semantic latent space instead of a high-dimensional data space leads to a notable reduction in computational costs due to reduced input size with a neglectable loss in synthesis quality \cite{rombach_high-resolution_2022}.
We construct a latent space by training an autoencoder with an encoder $E$ and decoder $D$. The autoencoder's task is to reconstruct \xlr, where the reconstruction is $D(E(\xlr))$ and $\z = E(\xlr) \in \mathcal{Z}$ represents a latent sample.
We follow \cite{rombach_high-resolution_2022, esser_taming_2021} for the autoencoder training and use a pixel-wise reconstruction loss next to a perceptual loss \cite{zhang_unreasonable_2018} and an adversarial objective.
\paragraph{Semantic Generation}

The actual synthesis diffusion model is trained by first obtaining $\z = E(\xlr)$ and modifying the optimization objective for the semantic generation as
\begin{equation} \label{eq:diff_on_latent}
    \mathbb{E}_{t\sim U(1,T), \bm z_0 \sim \mathcal{Z}, \bm \epsilon_t \sim \mathcal{N}(\bm 0, \bm 1)} \left[ \Vert \bm \epsilon_t - \bm \epsilon_{\theta}(\bm z_{t}, \tau_\phi(\bm y) ,t) \Vert^2_2 \right] \eqendp
\end{equation}
where $\bm \epsilon_\theta$ again is a time-conditional U-net. The diffusion process in \eqref{eq:diff_on_latent} is the same as described before, but operates only on samples in the latent space.
Additionally, we allow the conditioning on some modality $\bm y$ to guide the synthesis process. $\bm y$ can represent information of various kinds, e.g., radiological reports, class conditioning, or annotations. This information is fed into the model using a conditioning embedding, which is obtained by $\tau_\phi(\bm y)$, where $\tau_\phi$ is a neural network tasked with creating an embedding for $\bm y$.

\paragraph{Super-resolution}

After synthesizing \z\ and reconstructing \xlr\ from it, we apply an additional iterative refinement procedure to not only counteract blurriness, which is induced by the autoencoder but also to advance to realistic resolution domains in clinical practice.
Inspired by SR3 \cite{saharia_image_2021}, we condition a diffusion model with \xlr\ to infer an optimal high-resolution output \xhr.
Thus, the training objective is:
\begin{equation}
    \mathbb{E}_{t\sim U(1,T), \x_{\text{\tiny HR}}\sim\mathcal{X}_{\text{\tiny HR}}, \bm \epsilon_t \sim \mathcal{N}(\bm 0, \bm 1)} \left[ \Vert \bm \epsilon_t - \bm \epsilon_{\psi}(\x_{\text{\tiny HR}, t}, \x_{\text{\tiny LR}}, t) \Vert^2_2 \right] \eqendc
\end{equation}
where $\bm \epsilon_{\psi}$ is a denoising time-conditional U-net with conditioning on \xlr.

\section{MaCheX: Massive Chest X-ray Dataset} \label{sec:data}

We present a large-scale, open-source, diverse collection of chest X-ray images using a common interface for a major selection of open datasets and unify them as \textit{\textbf{Ma}ssive \textbf{Che}st \textbf{X}-Ray Dataset}(\textit{MaCheX}).
With \textit{MaCheX} we provide the largest available openly accessible composition of chest X-ray data and hope to support fair and unbiased future research in the area.
The multi-centric setup of the \textit{MaCheX} collection encourages diversity with sources from across the world and aims to be less prone to local biases.
\begin{table}[t]
\centering
\caption{Overview of the components of \textit{MaCheX}.}
\resizebox{0.9\linewidth}{!}{%
\begin{tabular}{lrrrrrrr} 
\toprule
& & \multicolumn{3}{c}{\textbf{No. samples}} \\ \cmidrule{3-5}
\textbf{Dataset} & \textbf{No. patients} & \textbf{Train} & \textbf{Test} & \textbf{ $\in$ \textit{MaCheX}} & \textbf{\%} & \textbf{Origin} \\ 
\midrule 
ChestX-ray14 \cite{wang_chestx-ray8_2017} & 23,152 & 86,523 & 25,595 & 86,523 & 13,28\% & Bethesda, MD, USA\\
CheXpert \cite{irvin_chexpert_2019} & 65,240 & 223,414 & 235 & 191,027 & 29,32\% & Palo Alto, CA, USA \\ 
MIMIC-CXR \cite{johnson_mimic-cxr_2019} & 65,379 & 377,110 & - & 243,334 & 37,35\% & Boston, MA, USA \\
PadChest \cite{bustos_padchest_2020} & 67,625 & 160,868 & - & 96,278 & 14,78\% & Alicante, Spain \\
BRAX \cite{reis_eduardo_pontes_brax_nodate}      & 18,529 & 40,967 & - & 19,309 & 2,96\% & São Paulo, Brazil \\
VinDr-CXR \cite{nguyen_vindr-cxr_2022} &  - & 15,000 & 3,000  & 15,000 & 2,30\% & Hanoi, Vietnam\\
\bottomrule
\end{tabular}
}
\label{tab:machex-overview}
\end{table}
Fostering a global data collection and increasing data fidelity is an important step toward offering a generalized solution without gender or racial prejudice.
Our multi-centric collection includes various datasets described in detail in the Supplementary Material and summarized in Table~\ref{tab:machex-overview}.

In the current version of \textit{MaCheX}, only frontal AP/PA scans are considered. By inclusion of the lateral position, another 250,000 chest X-rays could be added to the collection, totaling nearly a million samples. The present work, however, focuses on the analysis and synthesis of a frontal viewing position. 
All scans are rescaled so that the shortest edge meets a 1024px resolution and are then center-cropped to 1024 $\times$ 1024px.
We do not apply histogram equalization or other standardization techniques.
By combining the designated train subsets of the respective datasets, the full \textit{MaCheX} dataset amounts to 651,471 samples.
This includes over 440,000 labels, over 220,000 free-text radiological reports, and 15,000 coordinate bounding boxes for radiologist annotations.

We open-source our implementation of the pre-processing setup and provide an easy-to-use interface to access \textit{MaCheX} in a deep learning setting with PyTorch.
The structure of \textit{MaCheX} is clear and simple, allowing to straightforwardly adapt the code to different frameworks and use cases. 

\section{Experiments} \label{sec:experiment}

Our model stack effectively contains three models:
The semantic diffusion model (SDM), the autoencoder (AE), and the super-resolution diffusion model (SR).
AE and SR form a foundational basis that is trained on full \textit{MaCheX} with a separate test set of size 25,000. A full description of our training routine and hyperparameter setup can be found in the supplementary material.
For AE we use a KL-regularized continuous variational AE (VAE) with a downsampling factor 4 as in \cite{rombach_high-resolution_2022}.
In SR, we found that conditioning on bicubic downsampled versions of \xhr\ as \xlr\ as in \cite{saharia_image_2021} does not perfectly reflect the structure of recovered latent samples $D(\z)$ and results in artifacts and slight blurriness.
To align SR with the synthesis pipeline, we fine-tune SR on $D(\z)$.

SDM in the lower-dimensional (latent sample) domain is trained with a task-specific data subset of \textit{MaCheX}. 
To guide the synthesis process for chosen conditionings, we induce the respective conditioning embedding $\tau_\phi(\bm y)$ into the model over cross-attention \cite{vaswani_attention_2017} in the attention layers of the time-conditional U-net.

\paragraph{Reconstruction Quality}

\begin{figure}[t]
    \centering
    \begin{tabular}{cccc}
     \includegraphics[width=0.24\textwidth]{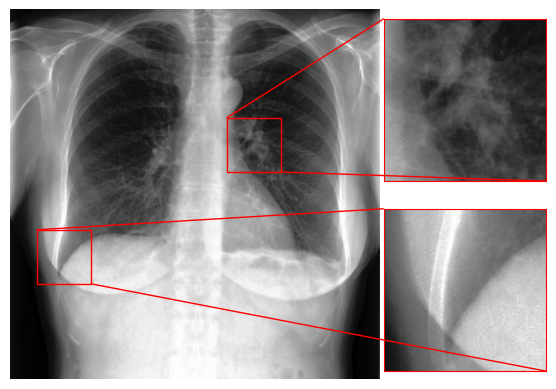} &   \includegraphics[width=0.24\textwidth]{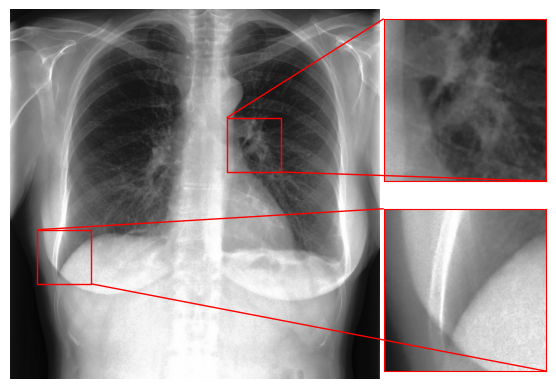} &
  \includegraphics[width=0.24\textwidth]{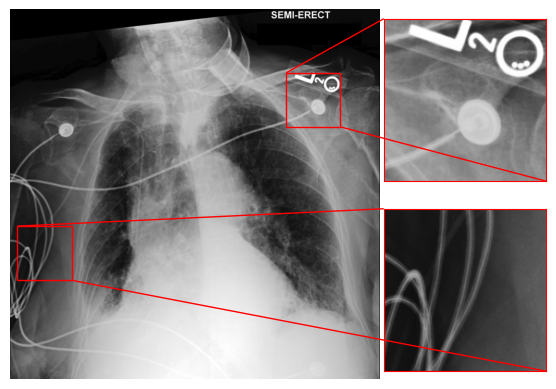} &   \includegraphics[width=0.24\textwidth]{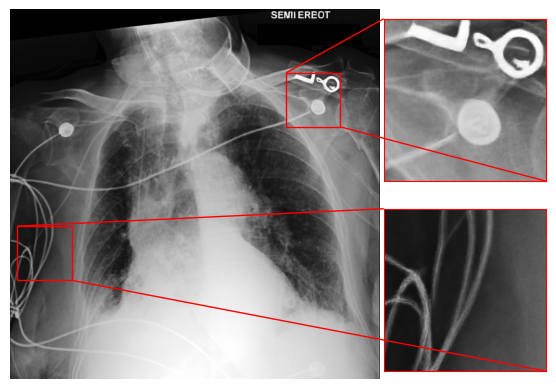} \\
    \end{tabular}
    \caption{Original images (the respective \textbf{left} image) compared to their reconstructions (the respective \textbf{right} image) from the latent space $\mathcal{Z}$. }
    \label{fig:ex-reconstruction}
\end{figure}

Using our AE and SR, a high-resolution image can be converted to the latent space and successfully reconstructed (cf.~Figure~\ref{fig:ex-reconstruction}). Despite a compression factor of $16$, even small details are recovered in fine-grained quality and limited blurriness, some structures even gaining sharpness.
Reconstruction of text annotations and small numbers, however, still proves to be difficult.
Table~\ref{tab:rec-metrics} additionally provides a quantitative assessment, i.e., the efficacy of the various transitions in \textit{Cheff} as well as the effect of fine-tuning SR, by analyzing the mean squared error (MSE), the structural similarity index measure (SSIM), and the peak signal-to-noise ratio (PSNR) on the test data set. 
The fine-tuned SR performs better than the base SR in almost every aspect, even in naive upscaling, while all transitions show high-quality reconstructions in general.
\begin{table}[t]
    \centering
        \caption{Test set performance for different reconstruction workflows. {\small \textcolor{black!75}{\textcircled{\raisebox{-0.9pt}{1}}}} symbolizes the encoding of \xhr\ and retrieval from \z\ in the full \textit{Cheff} pipeline, whereas {\small\textcolor{black!75}{\textcircled{\raisebox{-0.9pt}{2}}}} and {\small\textcolor{black!75}{\textcircled{\raisebox{-0.9pt}{3}}}} analyze the reconstruction capacity of AE and SR separately.
    }
    \vspace{2mm}
    \begin{tabular}{ll|rrr}
        \toprule
         \multicolumn{2}{l}{\textbf{Reconstruction workflow}} & \textbf{MSE} $\bm \downarrow$ & \textbf{PSNR} $\bm \uparrow$ & \textbf{SSIM} $\bm \uparrow$ \\
        \midrule

        \multirow{2}{*}{\textcolor{black!75}{\textcircled{\raisebox{-0.9pt}{1}}} \,} & $\xhr \xrightarrow[]{\text{bicubic}} \xlr \xrightarrow[]{E} \z \xrightarrow[]{D} \hat{\bm{x}}_{\text{\tiny LR}} \xrightarrow[]{\text{\textcolor{cb-green-sea}{SR}}_{\text{\textcolor{cb-green-sea}{base}}}}
        \hat{\bm{x}}_{\text{\tiny HR}}$   & 0.0039  & 24.05 & \textbf{0.9512} \\

        & $\xhr \xrightarrow[]{\text{bicubic}} \xlr \xrightarrow[]{E} \z \xrightarrow[]{D} \hat{\bm{x}}_{\text{\tiny LR}} \xrightarrow[]{\text{\textcolor{cb-burgundy}{SR}}_{\text{\textcolor{cb-burgundy}{fine}}}\, \,} \hat{\bm{x}}_{\text{\tiny HR}}$   & \textbf{0.0026}  & \textbf{25.77} & 0.9510 \\

        \arrayrulecolor{black!30}\midrule

        \textcolor{black!75}{\textcircled{\raisebox{-0.9pt}{2}}} \, & \phantom{$\xhr \xrightarrow[]{\text{bicubic}}$} $\xlr \xrightarrow[]{E} \z \xrightarrow[]{D} \hat{\bm{x}}_{\text{\tiny LR}}$   & 0.0005  & 36.80 & 0.9926 \\
        
        \midrule
        
        \multirow{2}{*}{\textcolor{black!75}{\textcircled{\raisebox{-0.9pt}{3}}} \,} & $\xhr \xrightarrow[]{\text{bicubic}} \xlr \xrightarrow[\hspace{28mm}]{\text{\textcolor{cb-green-sea}{SR}}_{\text{\textcolor{cb-green-sea}{base}}}}
        \hat{\bm{x}}_{\text{\tiny HR}}$   & 0.0035  & 24.59 & 0.9540 \\
        
        & $\xhr \xrightarrow[]{\text{bicubic}} \xlr 
        \xrightarrow[\hspace{28mm}]{\text{\textcolor{cb-burgundy}{SR}}_{\text{\textcolor{cb-burgundy}{fine}}}\, \,}
        \hat{\bm{x}}_{\text{\tiny HR}}$   & \textbf{0.0025}  &\textbf{ 26.09} & \textbf{0.9559}
        \\   
        
        \arrayrulecolor{black}

        \bottomrule
    \end{tabular}
    \label{tab:rec-metrics}
\end{table}

\paragraph{Unconditional Synthesis}

\begin{figure}[t]
     \centering
    \begin{subfigure}[b]{0.31\textwidth}
         \centering
    \begin{tabular}{cc}
  \includegraphics[width=0.48\textwidth]{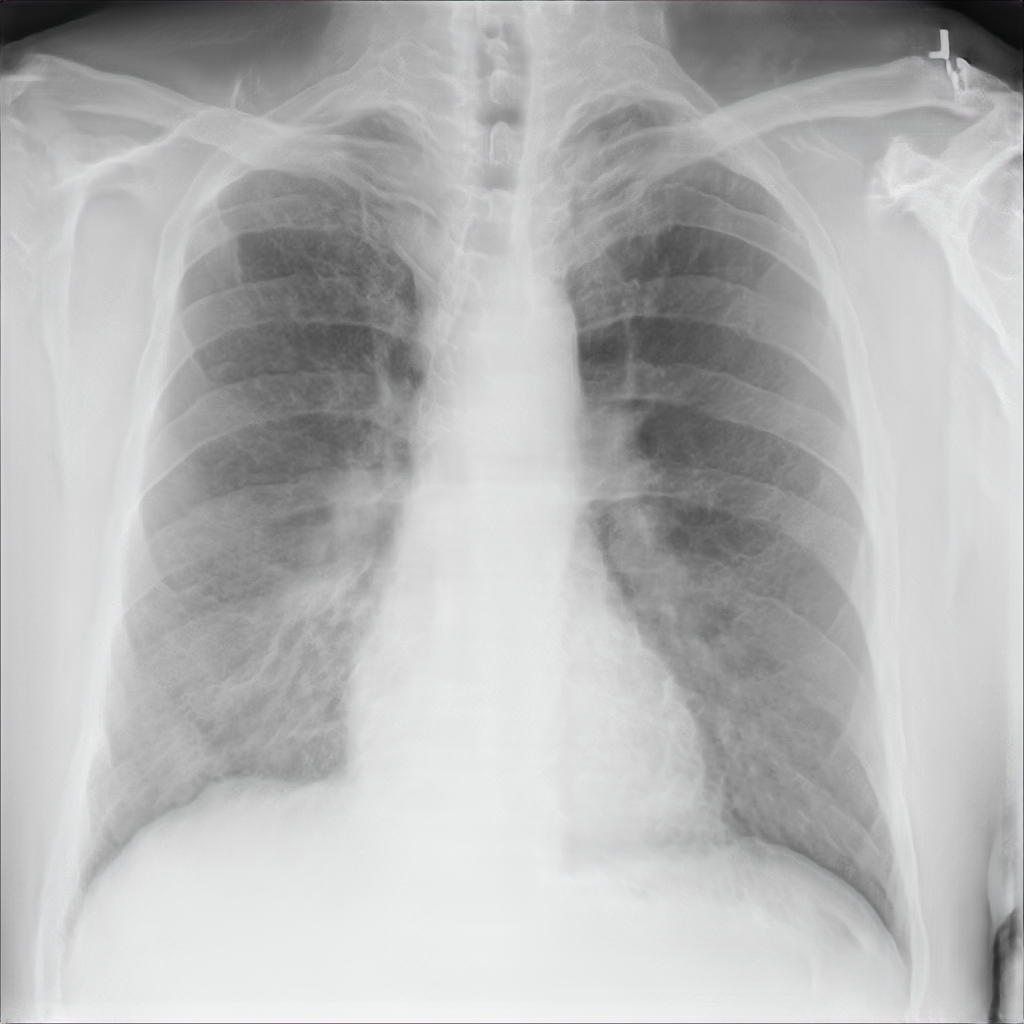} & 
  \includegraphics[width=0.48\textwidth]{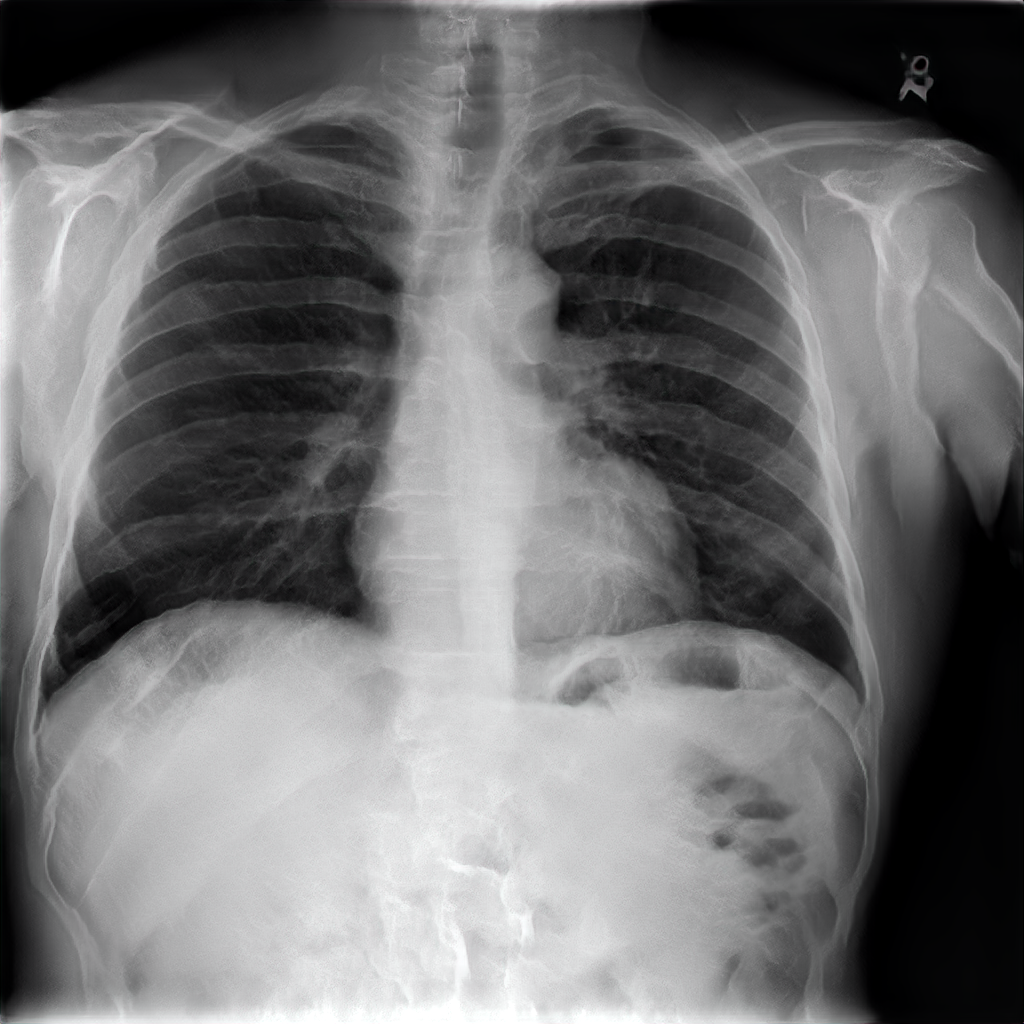} \\
    \end{tabular}
    \caption{PGAN \cite{segal_evaluating_2021}}
    \label{fig:ex-synth-segal}
     \end{subfigure}
    \hfill
    \begin{subfigure}[b]{0.31\textwidth}
         \centering
    \begin{tabular}{cc}
  \includegraphics[width=0.48\textwidth]{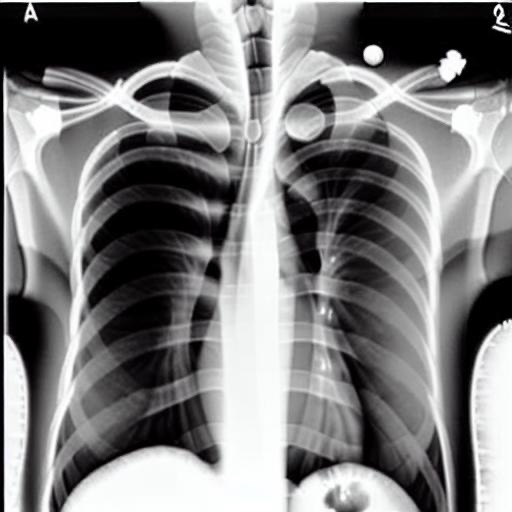} & 
  \includegraphics[width=0.48\textwidth]{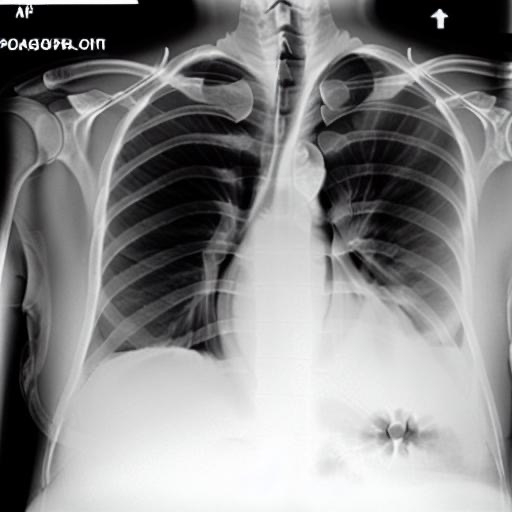} \\
    \end{tabular}
    \caption{Finetuned Stable Diff. \cite{chambon_adapting_2022} }
    \label{fig:ex-synth-chambon}
     \end{subfigure}
    \hfill
    \begin{subfigure}[b]{0.31\textwidth}
         \centering
    \begin{tabular}{cc}
  \includegraphics[width=0.48\textwidth]{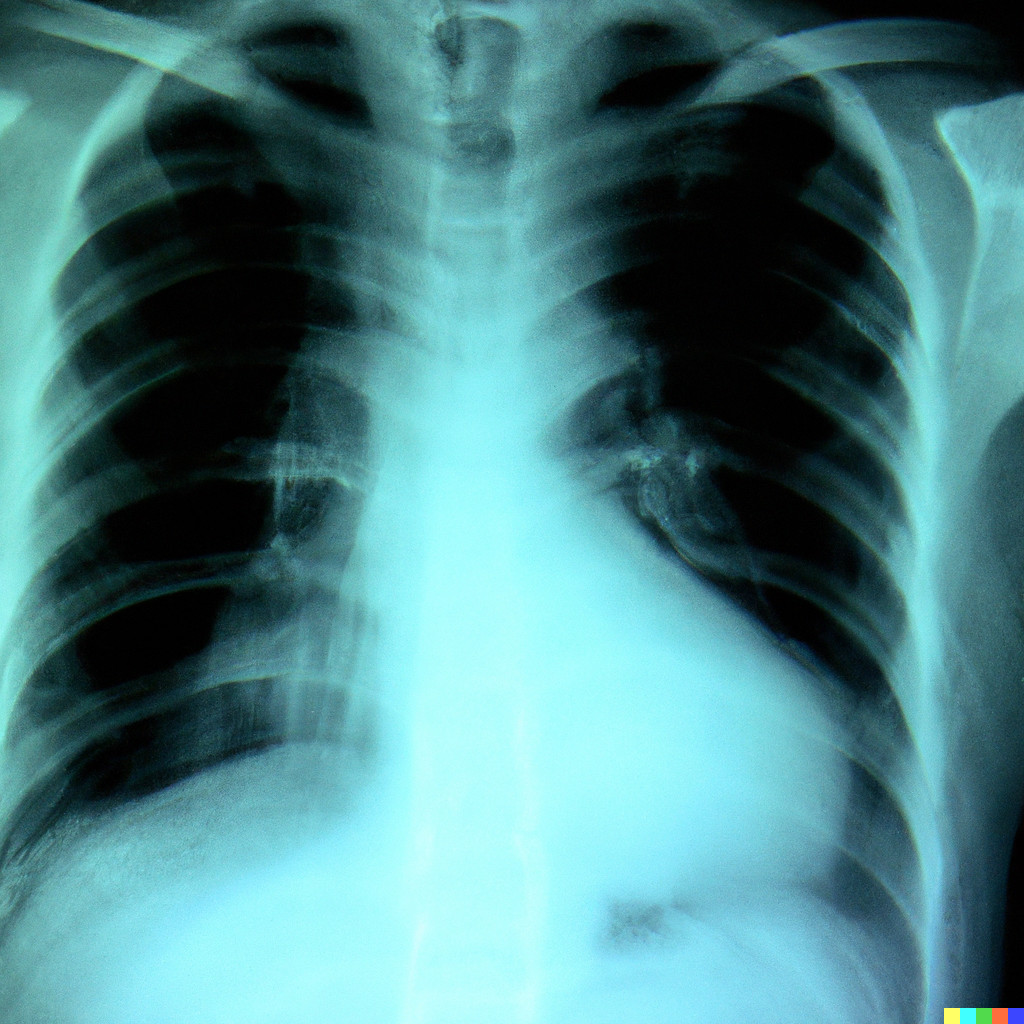} & 
  \includegraphics[width=0.48\textwidth]{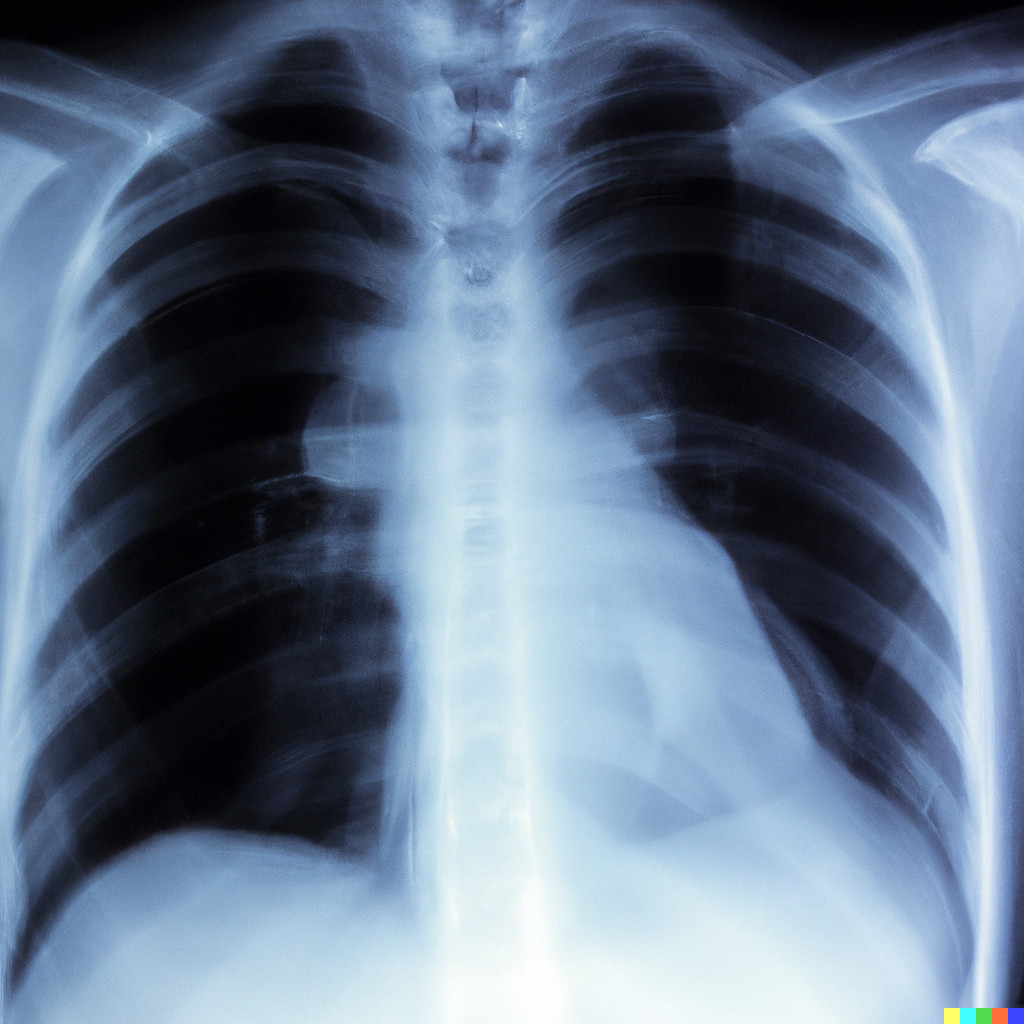} \\
    \end{tabular}
    \caption{Zero-shot DALL·E 2 \cite{ali_spot_2022}}
    \label{fig:ex-synth-ali}
     \end{subfigure}
    \par\medskip
     \begin{subfigure}[b]{\textwidth}
    \centering
    \begin{tabular}{cccccccc}
  \includegraphics[width=0.12\textwidth]{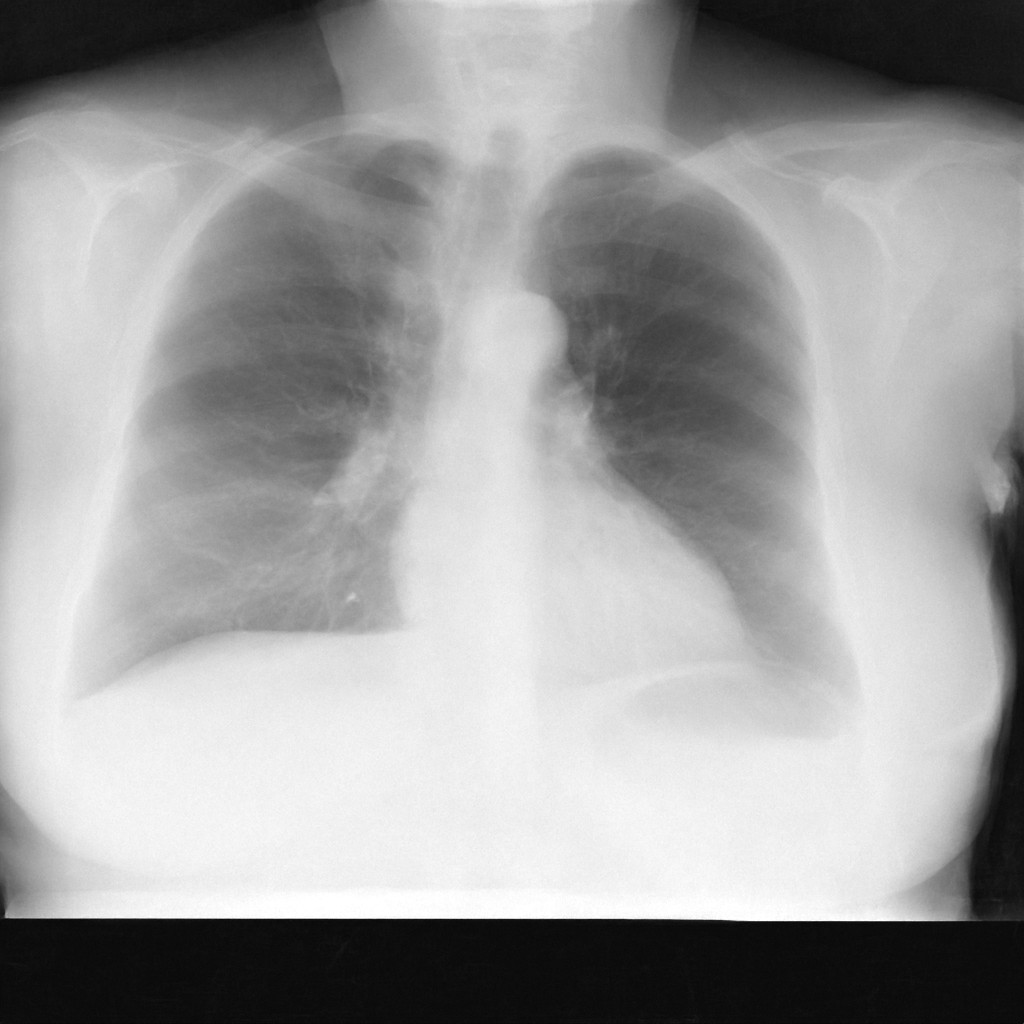} &
  \includegraphics[width=0.12\textwidth]{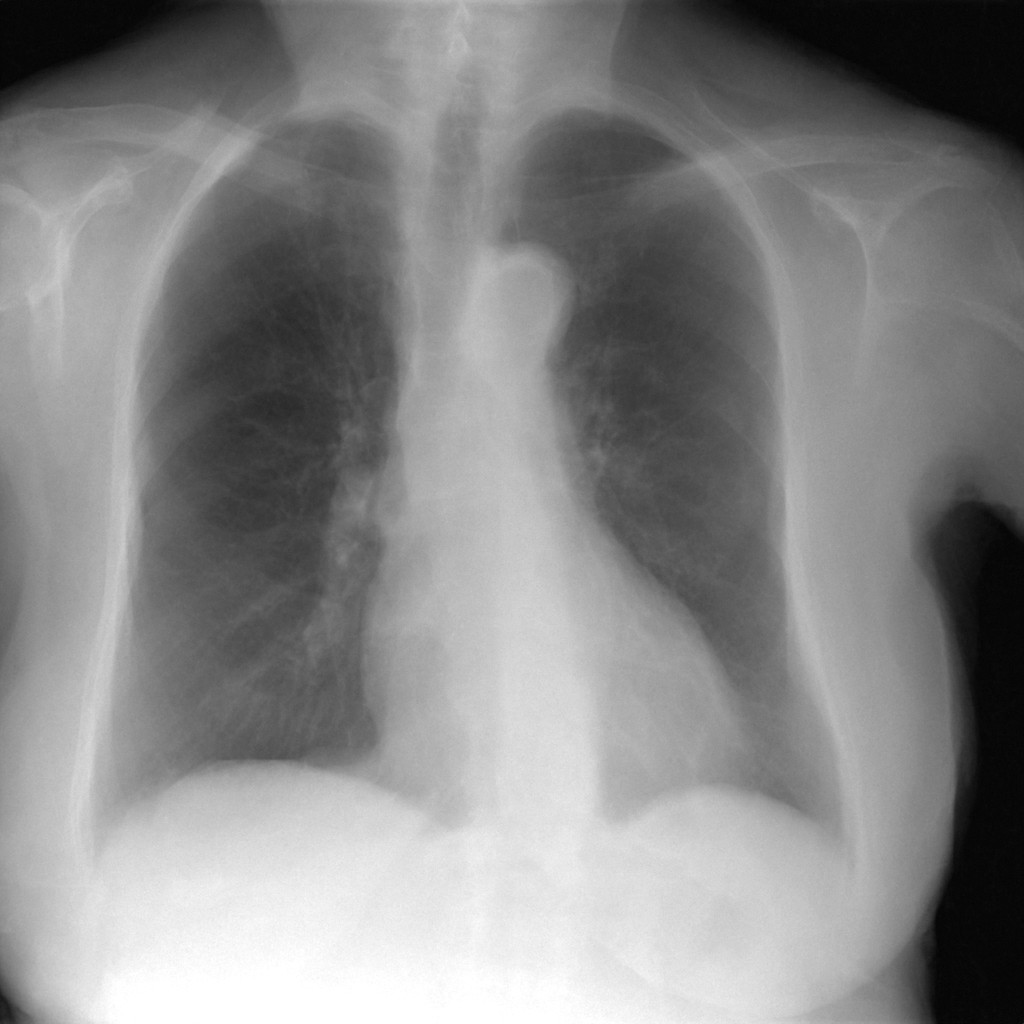} &
  \includegraphics[width=0.12\textwidth]{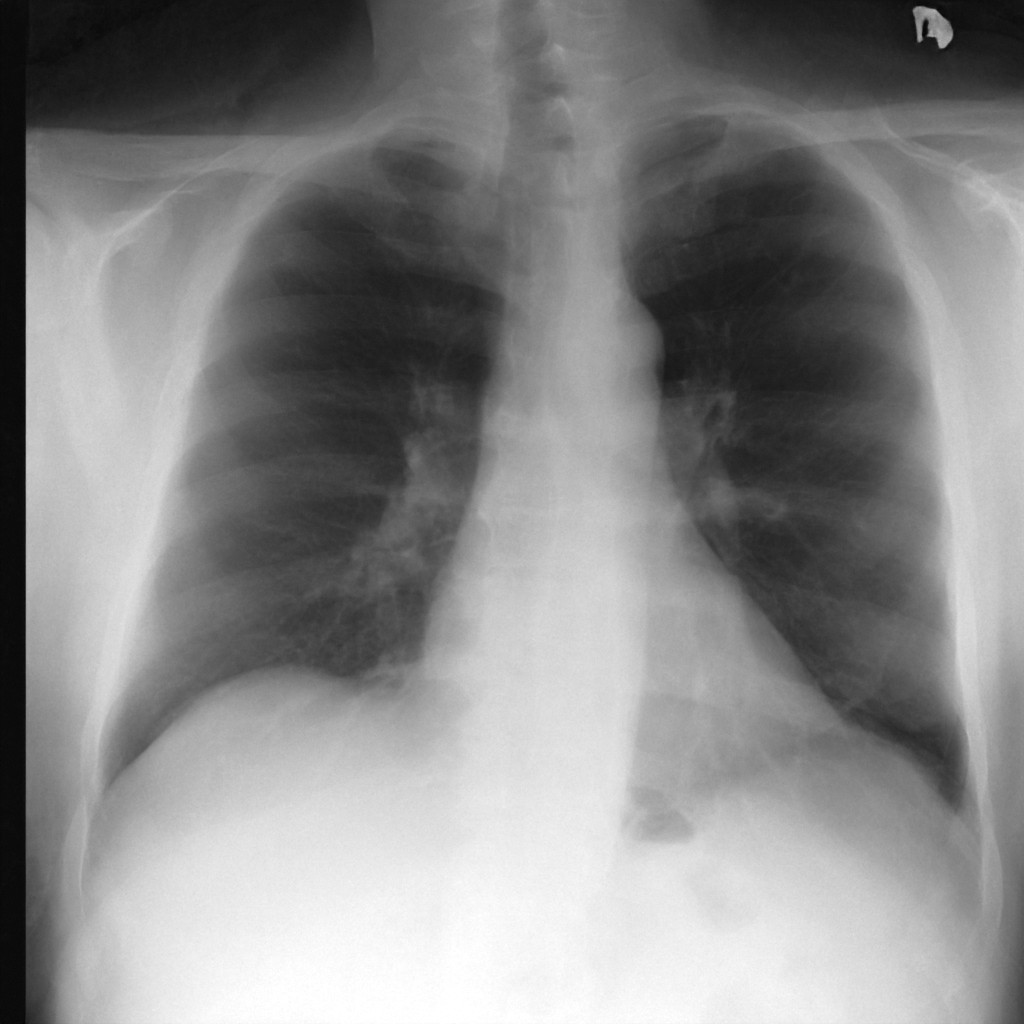} &
  \includegraphics[width=0.12\textwidth]{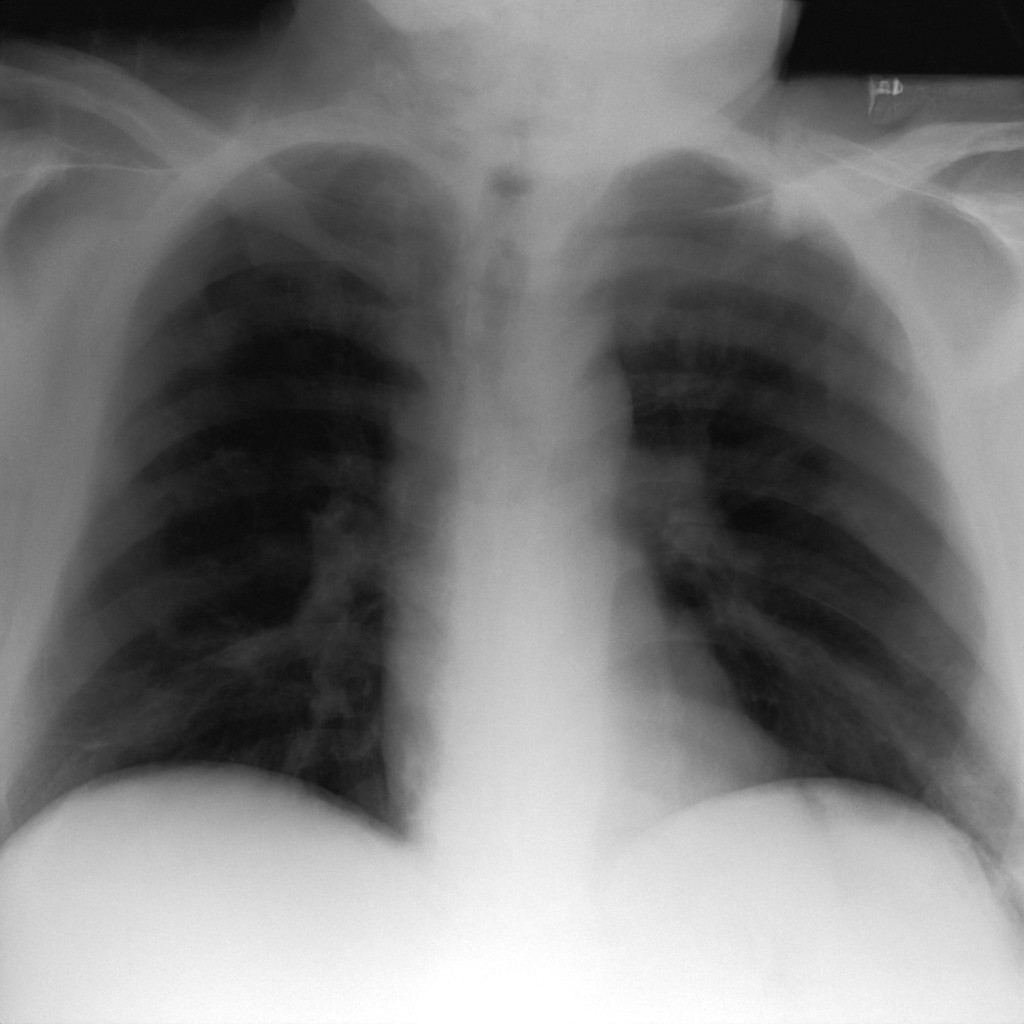} &
  \includegraphics[width=0.12\textwidth]{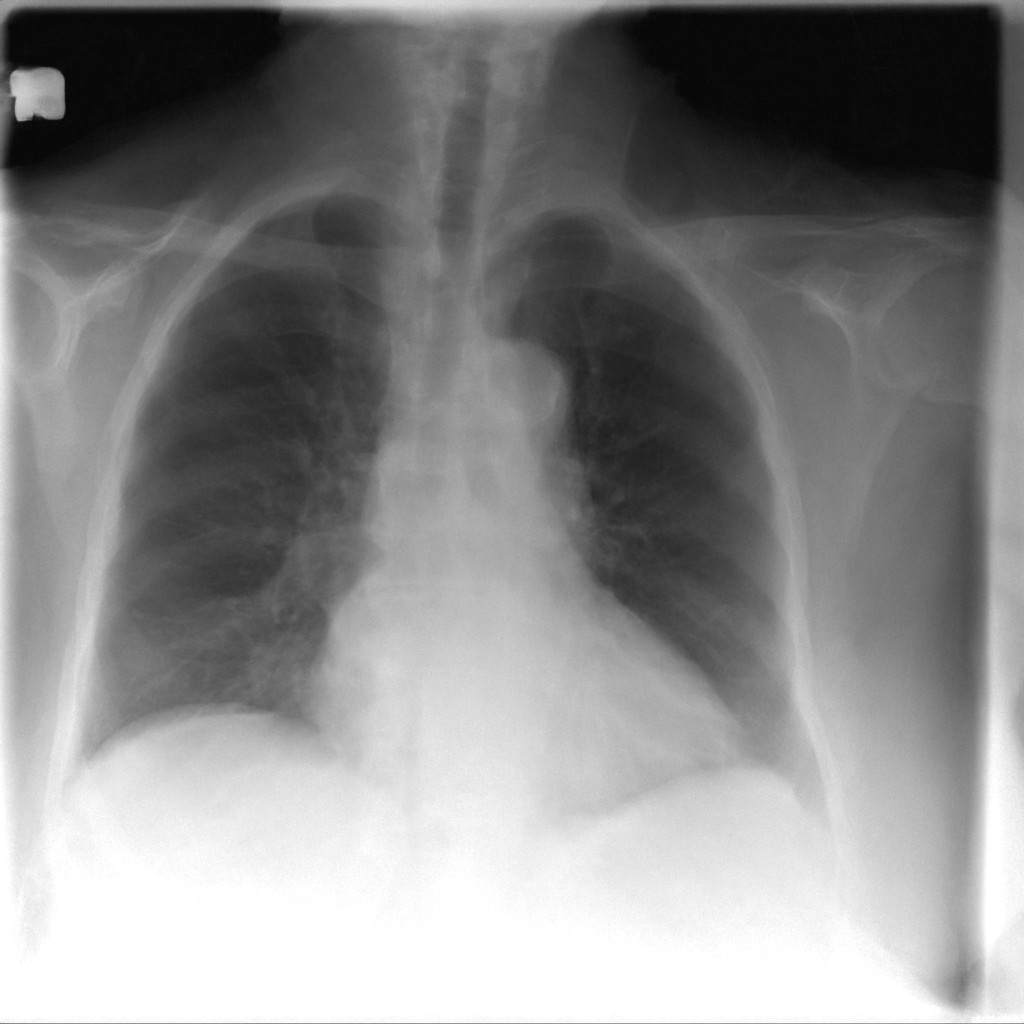} &
  \includegraphics[width=0.12\textwidth]{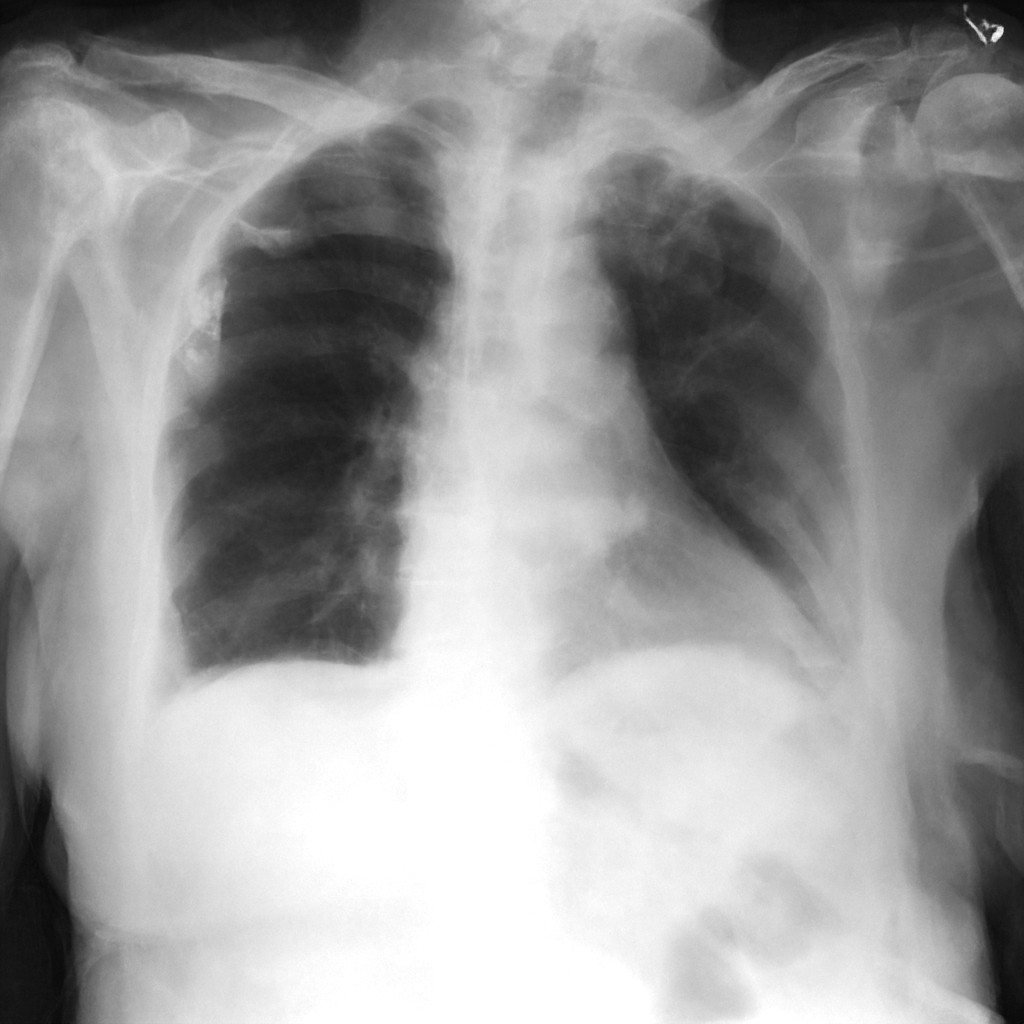} &
  \includegraphics[width=0.12\textwidth]{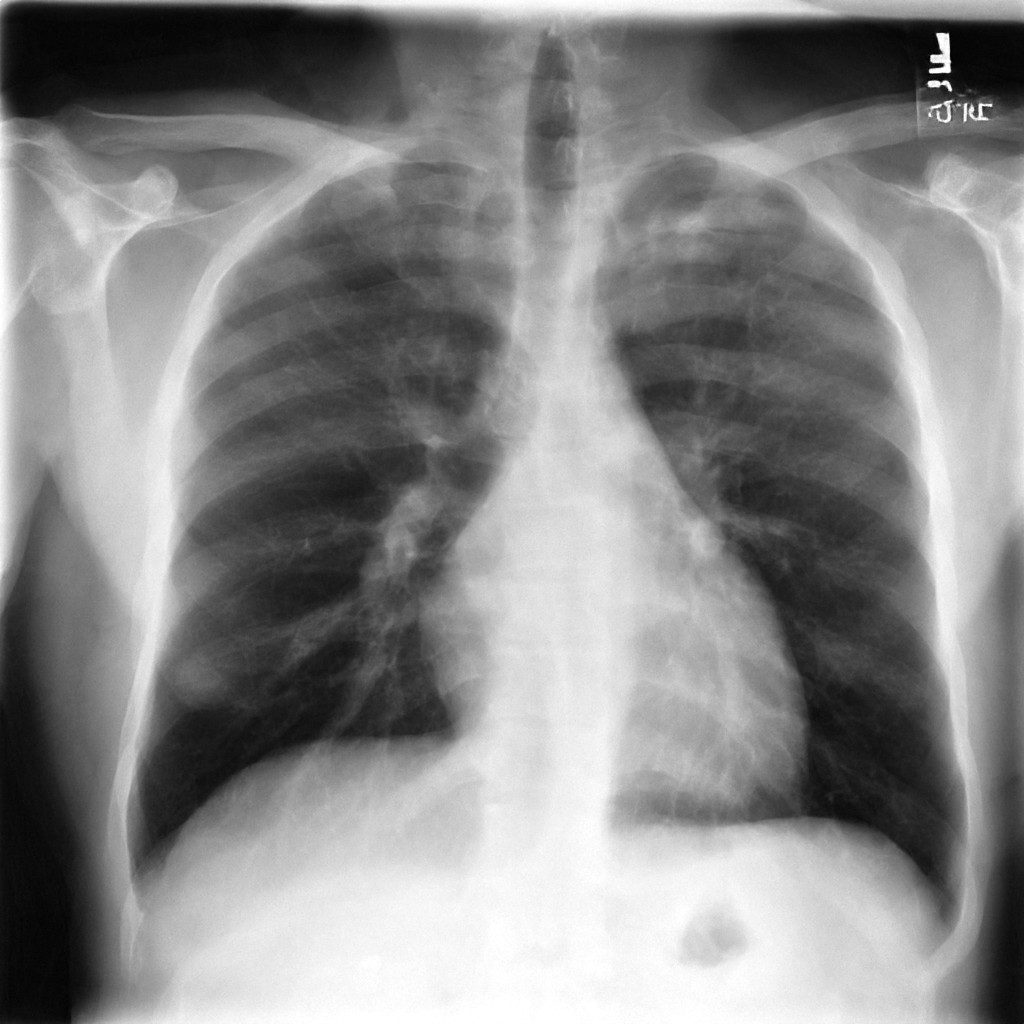} &
  \includegraphics[width=0.12\textwidth]{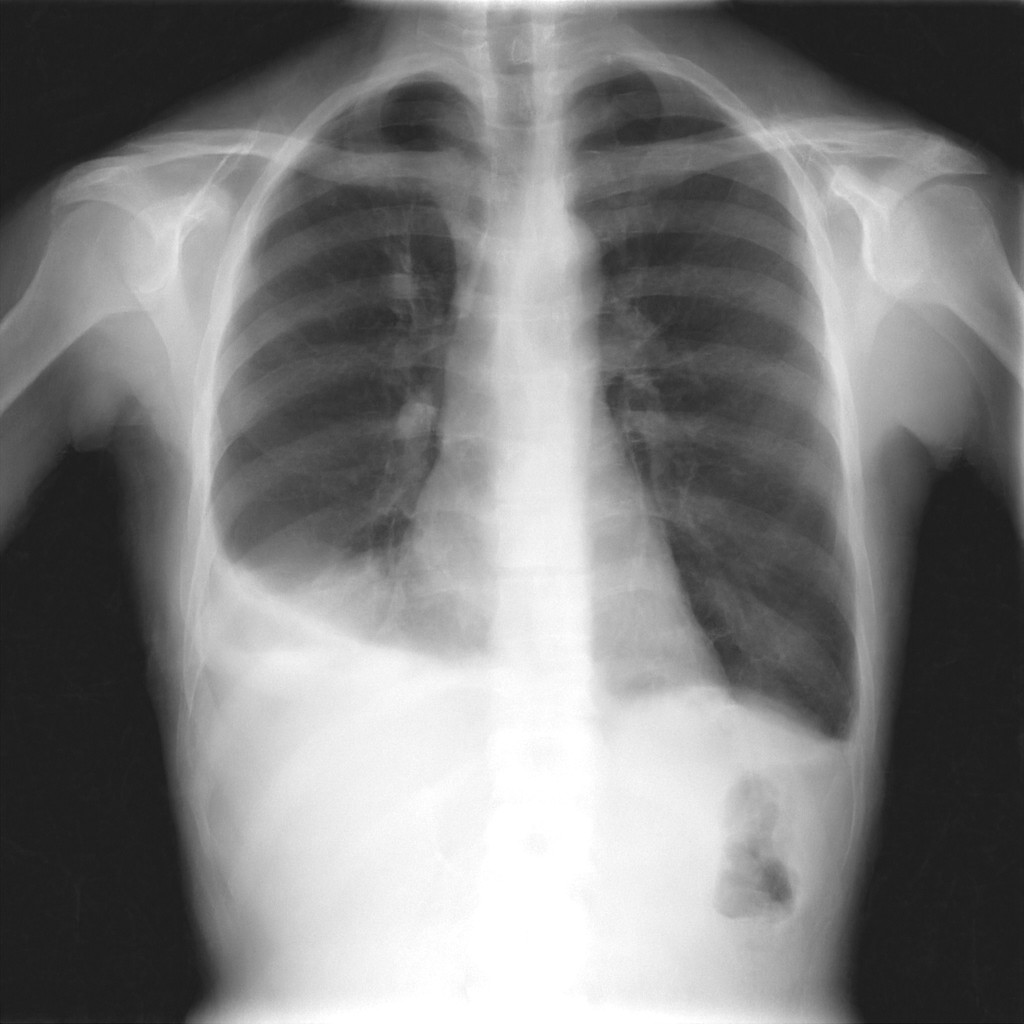} \\
    \end{tabular}
    \caption{Triple cascaded diffusion model.}
    \label{fig:ex-synth-casc}
     \end{subfigure}
          \par\medskip
\begin{subfigure}[b]{\textwidth}
    \centering
    \begin{tabular}{cccccccc}
  \includegraphics[width=0.12\textwidth]{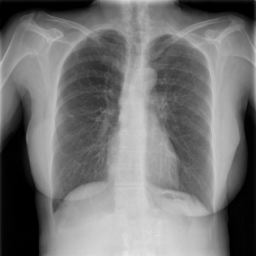} &
  \includegraphics[width=0.12\textwidth]{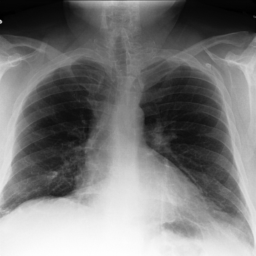} &
  \includegraphics[width=0.12\textwidth]{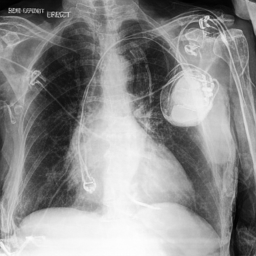} &
  \includegraphics[width=0.12\textwidth]{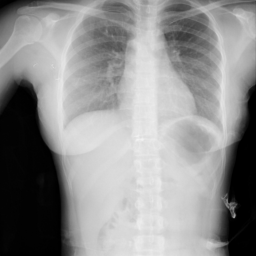} &
  \includegraphics[width=0.12\textwidth]{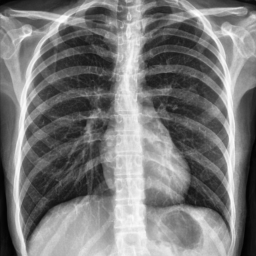} &
  \includegraphics[width=0.12\textwidth]{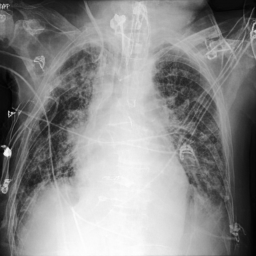} &
  \includegraphics[width=0.12\textwidth]{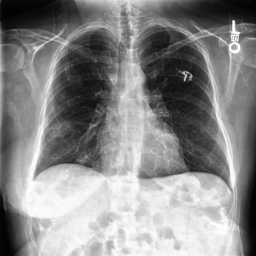} &
  \includegraphics[width=0.12\textwidth]{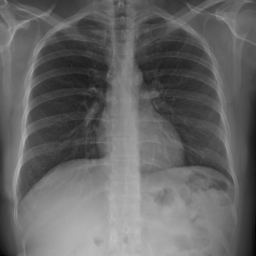} \\
  
  \includegraphics[width=0.12\textwidth]{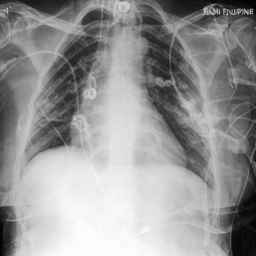} &
  \includegraphics[width=0.12\textwidth]{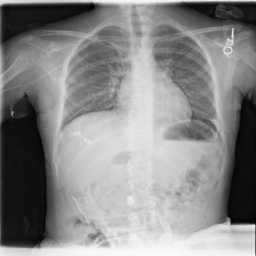} &
  \includegraphics[width=0.12\textwidth]{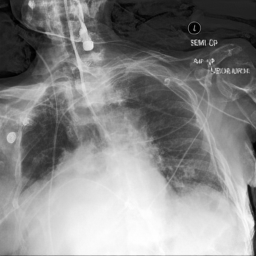} &
  \includegraphics[width=0.12\textwidth]{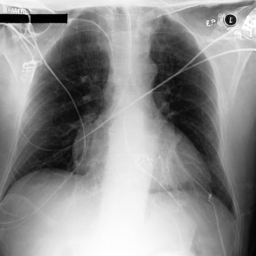} &
  \includegraphics[width=0.12\textwidth]{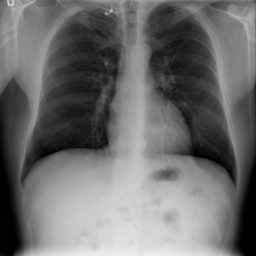} &
  \includegraphics[width=0.12\textwidth]{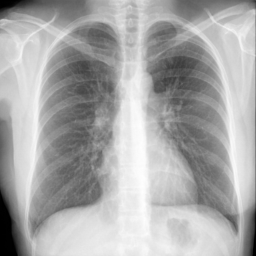} &
  \includegraphics[width=0.12\textwidth]{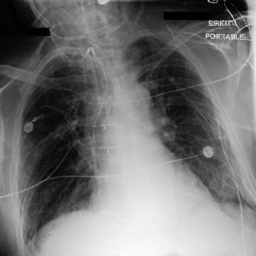} &
  \includegraphics[width=0.12\textwidth]{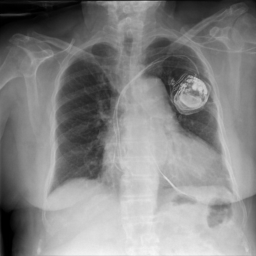} \\
    \end{tabular}
    \caption{\textit{Cheff} (Ours).}
    \label{fig:ex-synth-cheff}
     \end{subfigure}
     
    \caption{Synthesized chest X-rays using different methods.}
    \label{fig:ex-synth}
\end{figure}

Scaling up the data and model stack leads to a realistic synthesis with high fidelity and realism (Figure~\ref{fig:ex-synth-cheff}) in a 1-megapixel resolution setting including a variety of medical devices, e.g., chest tubes, pacemakers, ECG leads, etc.
This is a major improvement over previously proposed approaches for chest X-ray synthesis (Figure~\ref{fig:ex-synth}).
The PGAN of \cite{segal_evaluating_2021} misses the capacity of medical devices (cf. \cite{weber_implicit_2022}).
While synthetic samples of \cite{ali_spot_2022} by prompting DALL·E 2 are comparatively realistic for a non-medical zero-shot setting, they involve color jittering.
In contrast, Stable Diffusion is not able to generate clinically realistic chest X-rays, clearly showing its artistic style even when fine-tuned \cite{chambon_adapting_2022}.
LDM training of \cite{packhauser_generation_2022} involves only a 256-pixel resolution of a single data source.

%
In comparison to our latent diffusion-based pipeline, we investigate a fully cascaded stack of diffusion models in the style of Imagen \cite{saharia_photorealistic_2022} and DALL·E 2 \cite{ramesh_hierarchical_2022} with U-net configurations from \cite{saharia_image_2021} (see Figure \ref{fig:ex-synth-casc}).
The cascade involves three models in an upscaling chain: 64px $\rightarrow$ 256px $\rightarrow$ 1024px, where the first model synthesizes samples from noise.
This uncovers a potential downside of utilizing models with a progressive growing backend (same as in \cite{segal_evaluating_2021}). While the samples maintain high quality, there is also an abundance of medical devices and minor foreign objects.
This fact also reflects in the Fréchet Inception Distance (FID $\downarrow$) and Kernel Inception Distance (KID $\downarrow$), which is 46.54 and 0.0530 for the full cascade but reaches 11.58 and 0.0099, respectively, using \textit{Cheff}. 
We hypothesize that details of this granularity are lost when synthesizing on the low-resolution levels of $\leq$ 64 pixels.
The subsequent upsampling models cannot recover these already disregarded elements.
Our approach works on semantic features instead of low-resolution images and thus is able to compress necessary information in a meaningful way, successfully circumventing this issue (Figure \ref{fig:ex-synth-cheff}).
%

\begin{figure}[t]
    \centering
    \begin{tabular}{cccc}
     \includegraphics[width=0.24\textwidth]{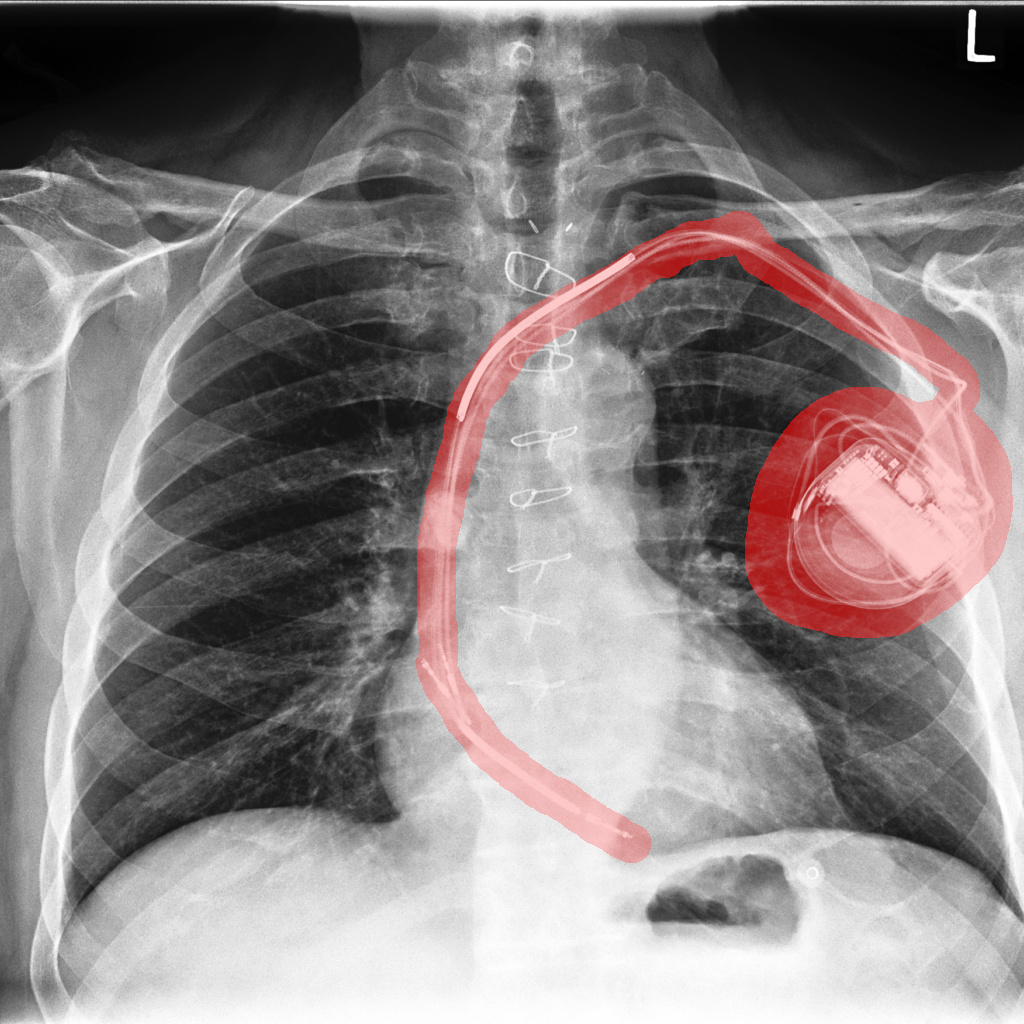} &   \includegraphics[width=0.24\textwidth]{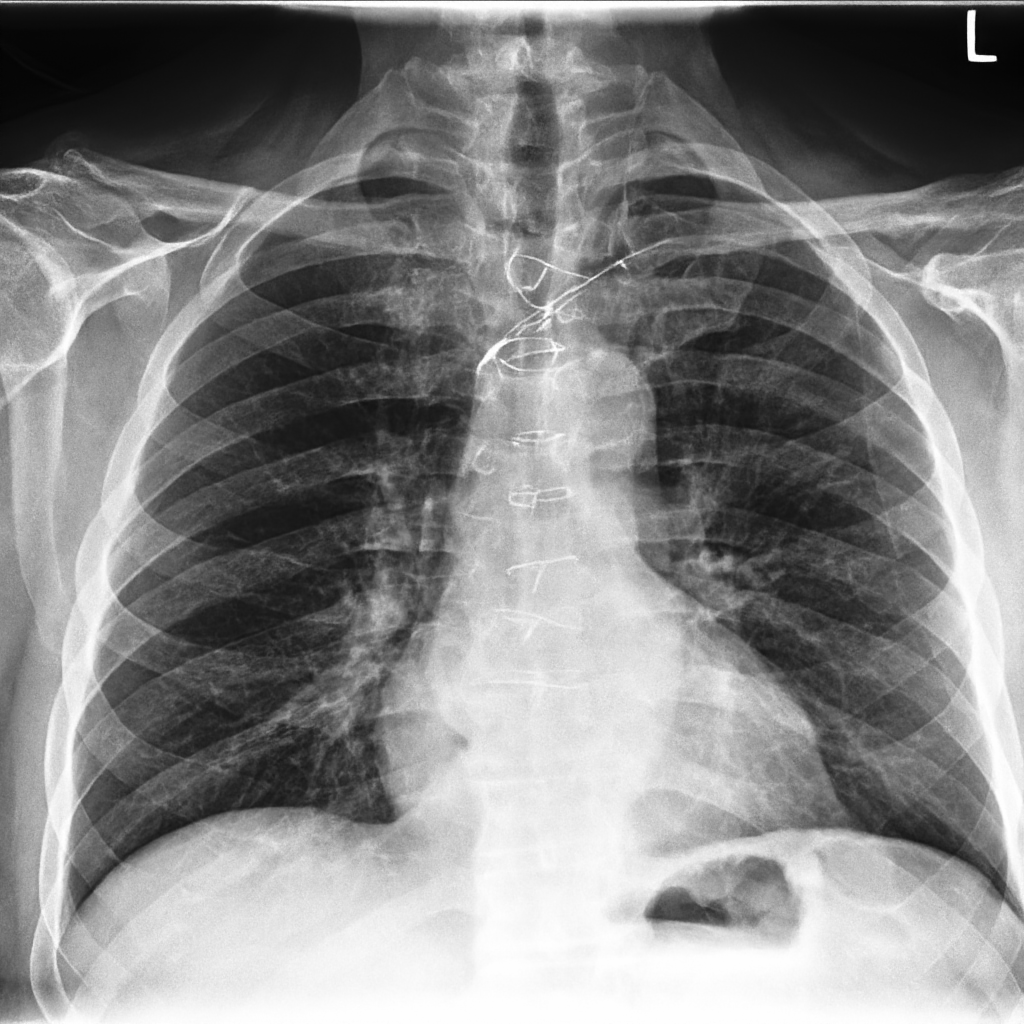} &
  \includegraphics[width=0.24\textwidth]{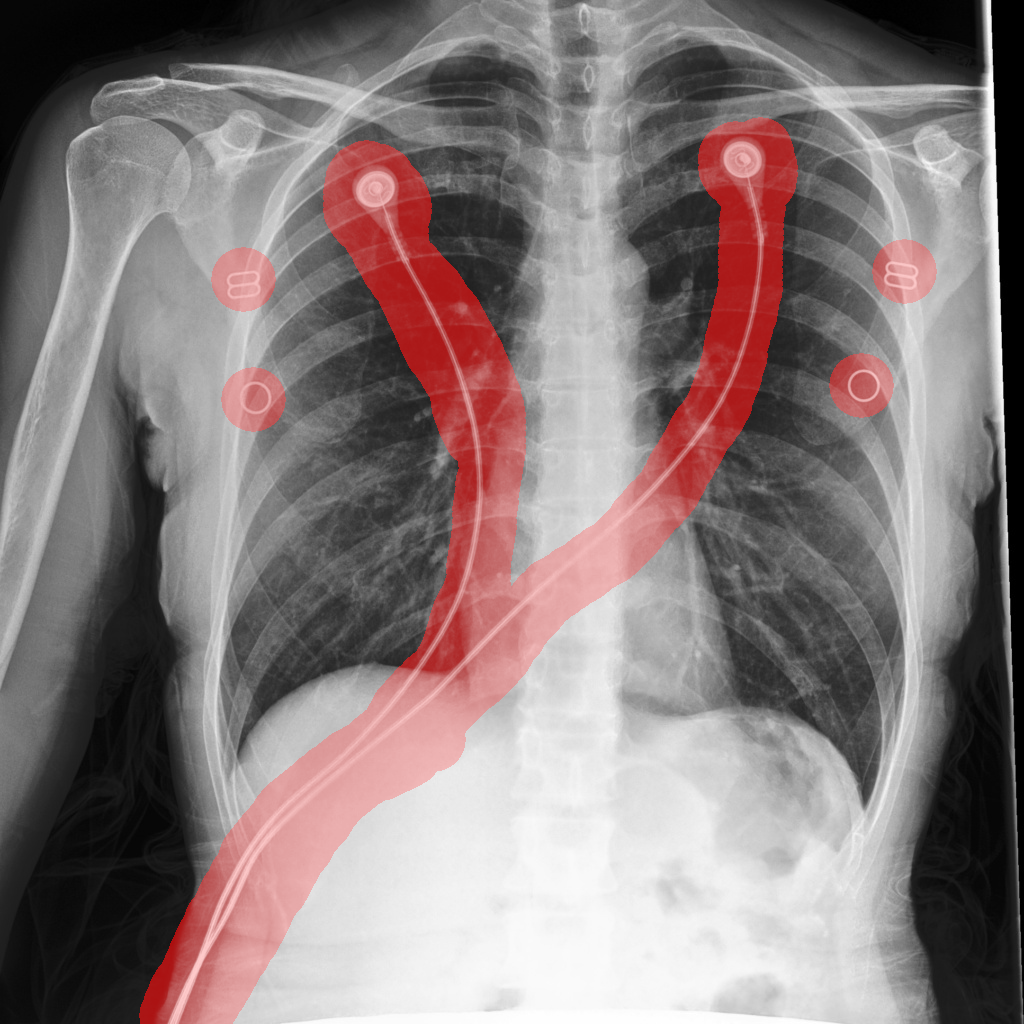} &   \includegraphics[width=0.24\textwidth]{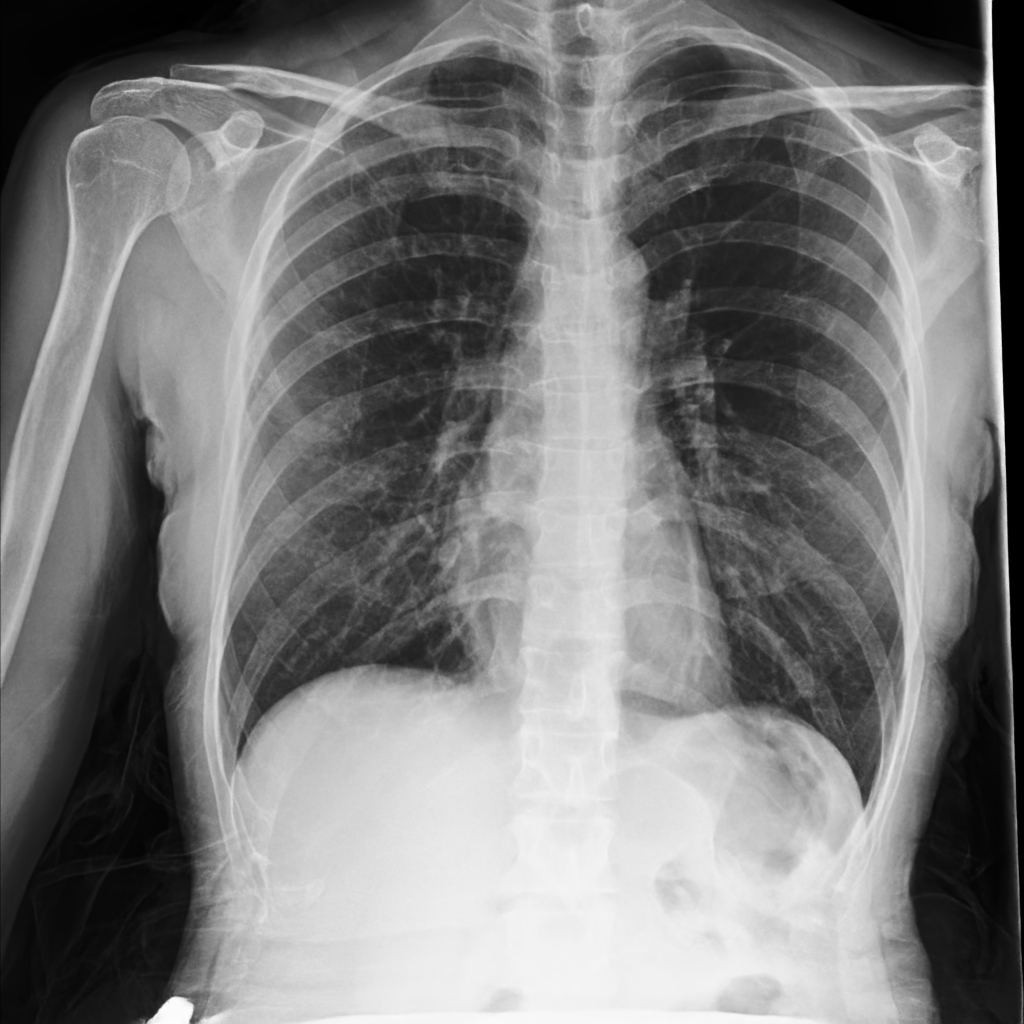} \\
    \end{tabular}
    \caption{Removal of support devices via image inpainting.}
    \label{fig:ex-inpaint}
\end{figure}
\begin{figure}[t]
\centering
\begin{minipage}[t]{.33\textwidth}
  \centering
  \begin{tabular}{cc}
     \includegraphics[width=0.5\textwidth]{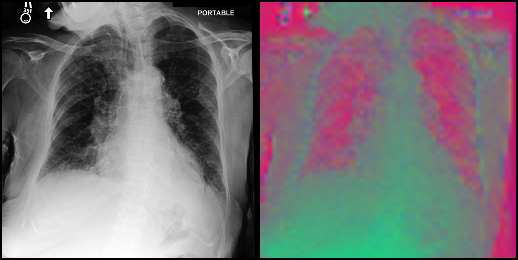} &   \includegraphics[width=0.5\textwidth]{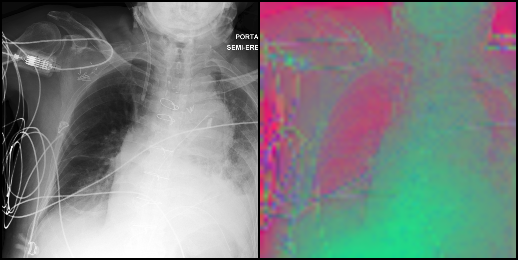} \\
          \includegraphics[width=0.5\textwidth]{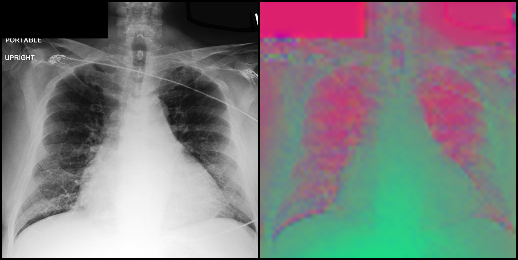} &   \includegraphics[width=0.5\textwidth]{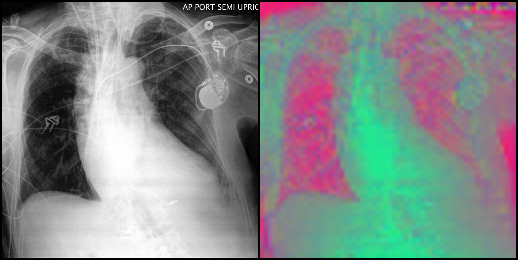} \\
     \includegraphics[width=0.5\textwidth]{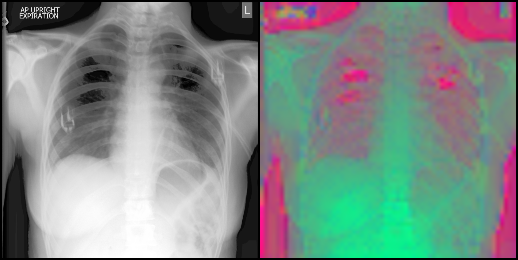} &   \includegraphics[width=0.5\textwidth]{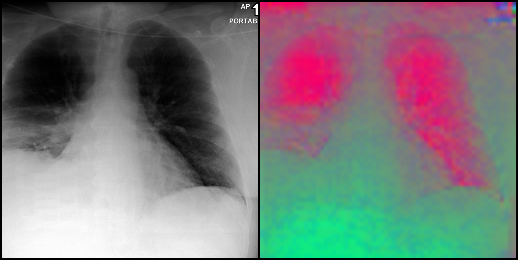} \\
    \end{tabular}
  \captionof{figure}{Comparison of original image (\textbf{left}) with latent embedding (\textbf{right}) shows spatial coherency.}
  \label{fig:ex-spatial}
\end{minipage}\hfill
\begin{minipage}[t]{.64\textwidth}
  \centering
  \begin{tabular}{cccc}
     \includegraphics[width=0.22\textwidth]{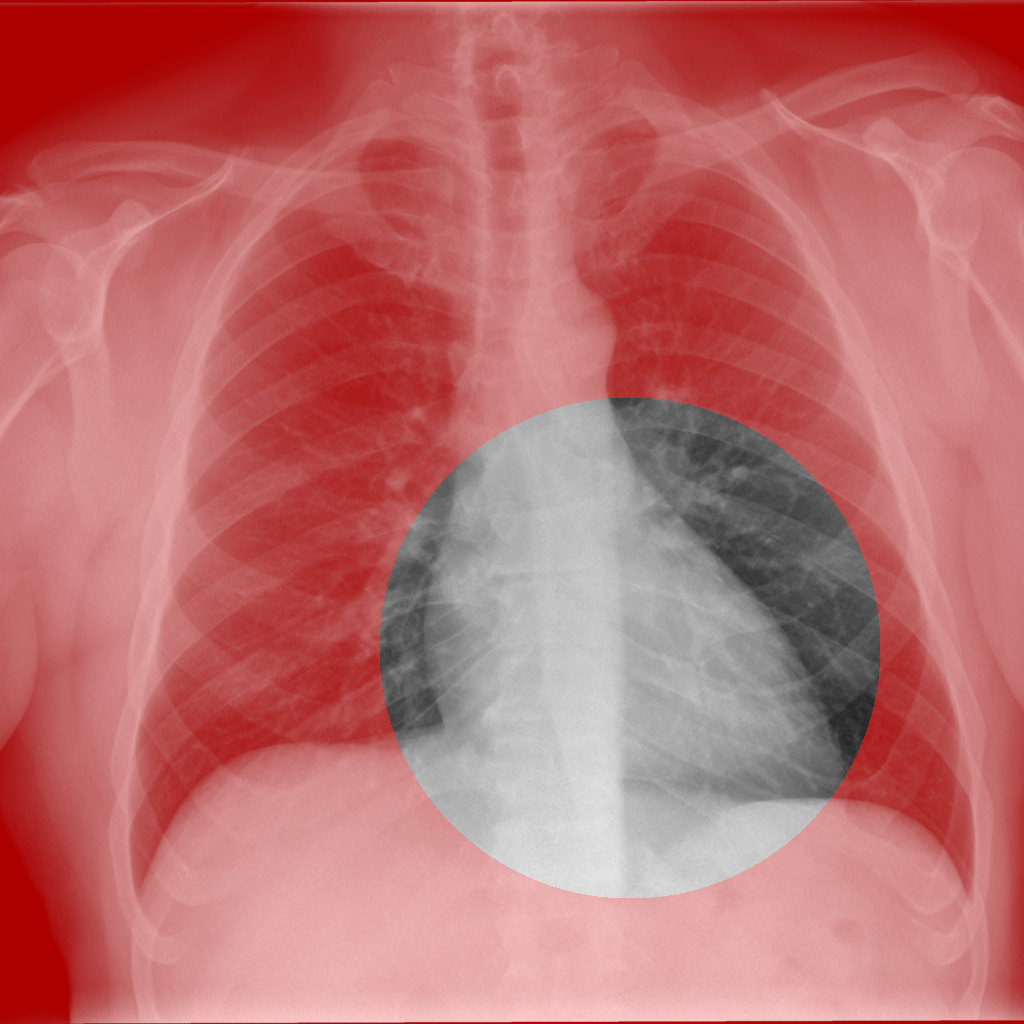} & \includegraphics[width=0.22\textwidth]{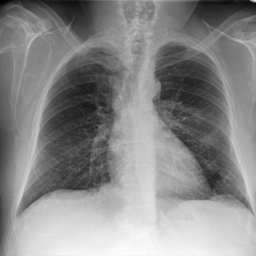} &  \includegraphics[width=0.22\textwidth]{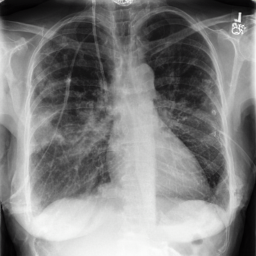} &  \includegraphics[width=0.22\textwidth]{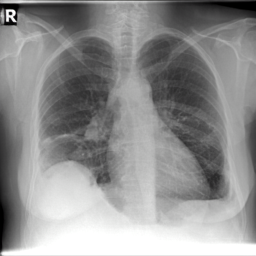} \\
     
     \includegraphics[width=0.22\textwidth]{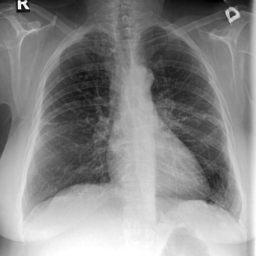} & \includegraphics[width=0.22\textwidth]{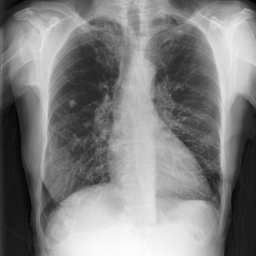} &  \includegraphics[width=0.22\textwidth]{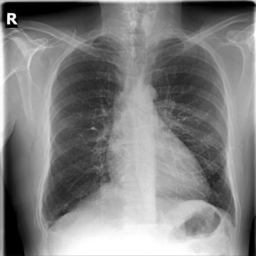} &  \includegraphics[width=0.22\textwidth]{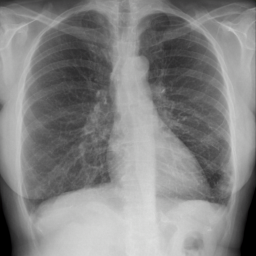} \\
    \end{tabular}
  \captionof{figure}{Outpainting variations of the highlighted area in the \textbf{upper left} image.}
  \label{fig:ex-outpaint}
\end{minipage}
\end{figure}

\paragraph{In- and Outpainting} The iterative nature of diffusion models allows for simple adaption of inpainting tasks, where the model is used to fill a designated marked area.
This method can be utilized to, e.g., remove occluding and distracting elements like medical devices from a thoracic scan (Figure~\ref{fig:ex-inpaint}).
While \textit{Cheff} has the ability to synthesize removed parts such as devices, object removal works well since  the rest of the image is provided and acts as a prior. Thus \textit{Cheff} fills the removed parts with the most likely content -- a clear lung -- and not the original devices.
In \textit{Cheff} or LDMs in general, this kind of masking is possible as the commonly used autoencoder implementation produces a spatially-aware convolutional embedding, which, despite living in a latent space, maintains structural information of the input data (Figure \ref{fig:ex-spatial}).
In another formulation, masking as outpainting serves to interpolate the remaining parts of an image. As seen in Figure~\ref{fig:ex-outpaint}, a variety of chest X-rays can be synthesized sharing the same initial provided area. Notably, \textit{Cheff} does not collapse to one solution but is able to explore various in-distribution options.

\begin{figure}[htbp]
     \centering
     \begin{subfigure}[b]{\textwidth}
         \centering
    \begin{tabular}{ccccc}
  \includegraphics[width=0.18\textwidth]{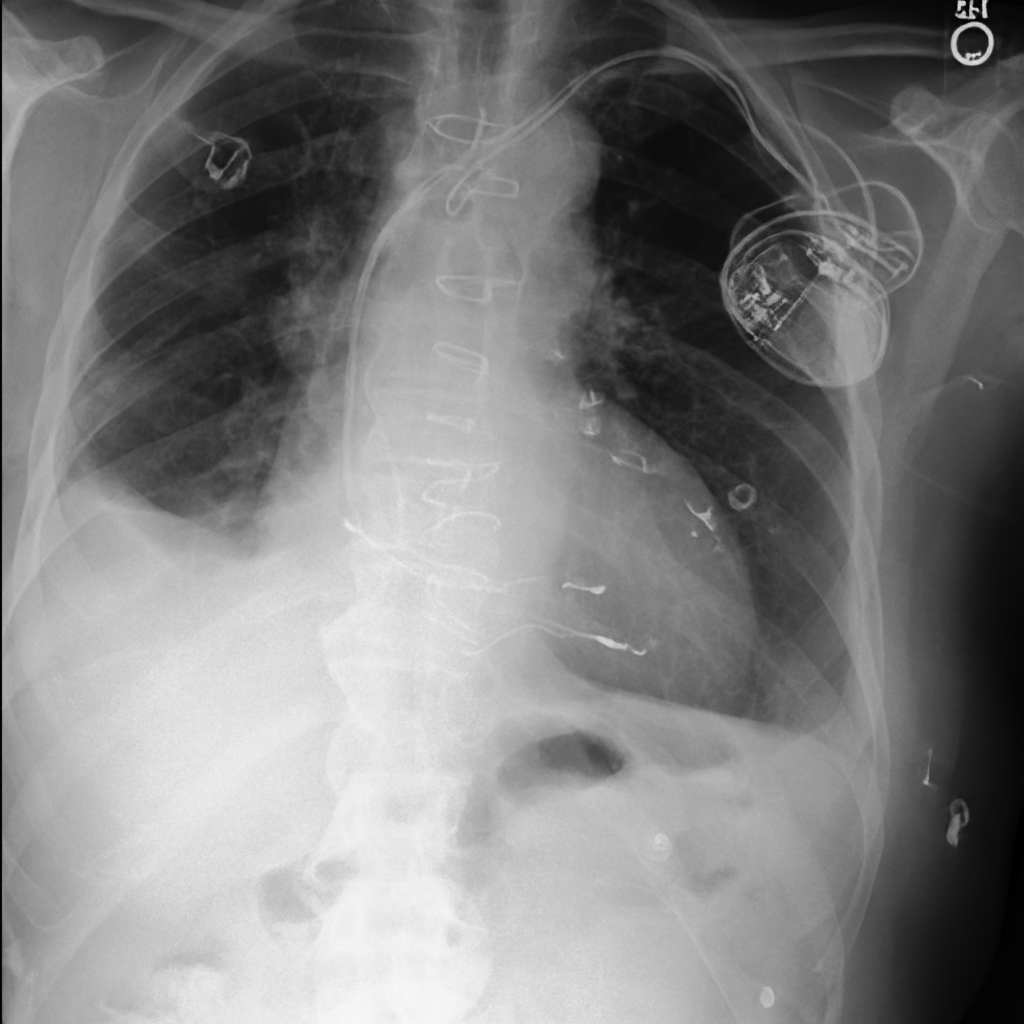} & \includegraphics[width=0.18\textwidth]{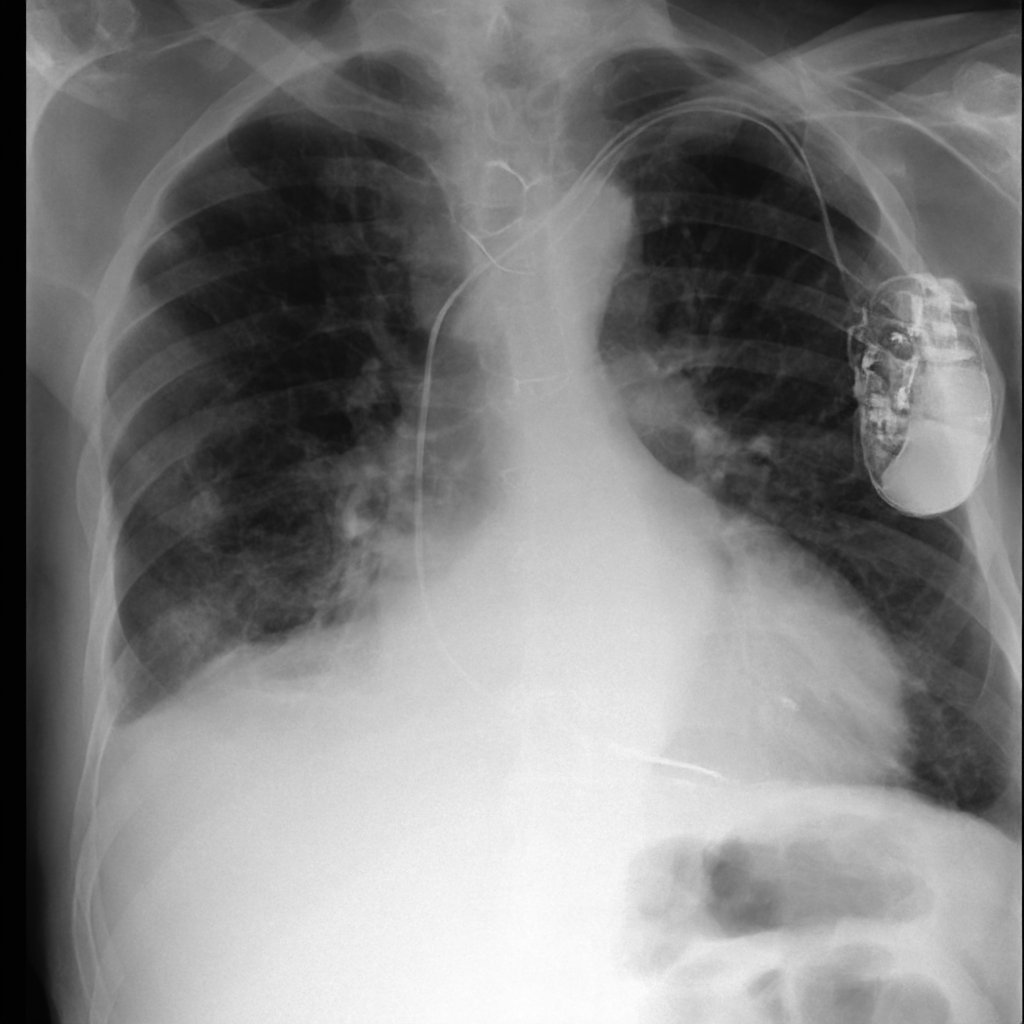} & \includegraphics[width=0.18\textwidth]{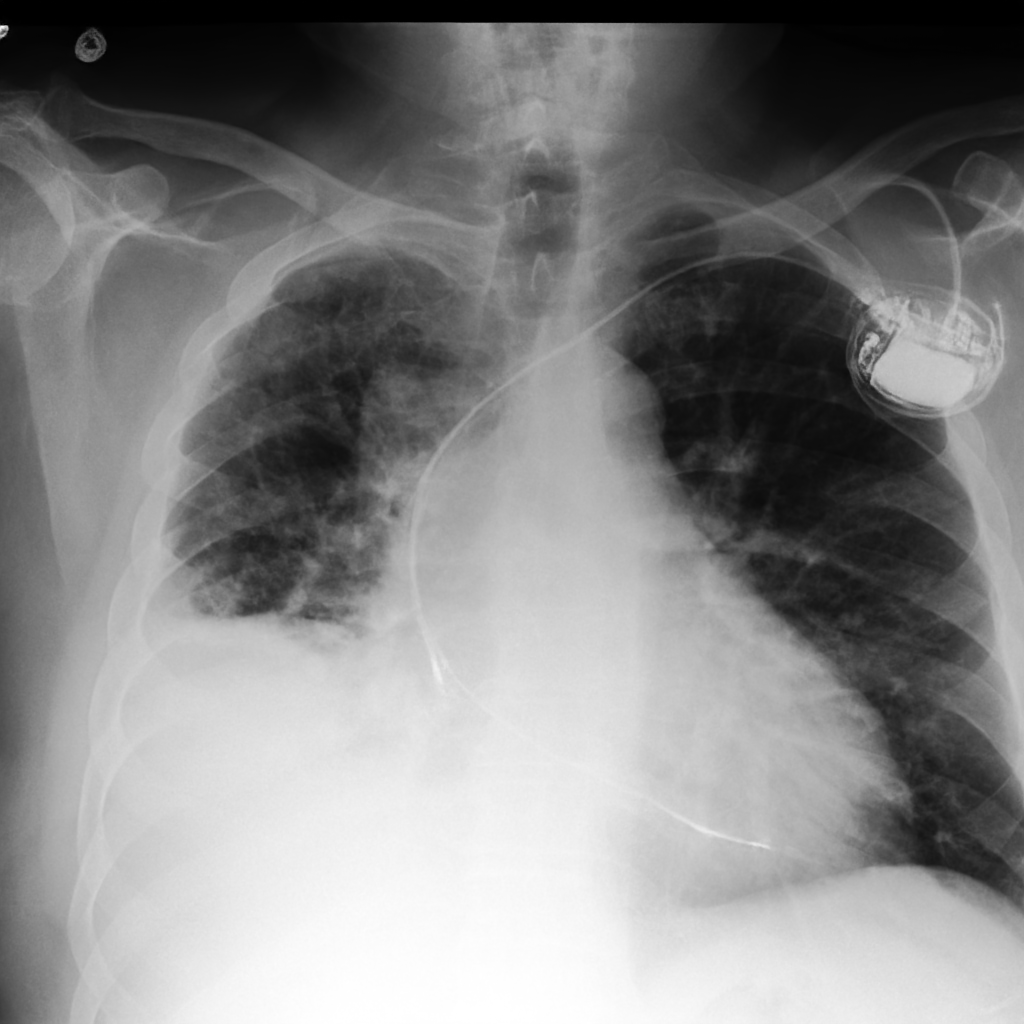} & \includegraphics[width=0.18\textwidth]{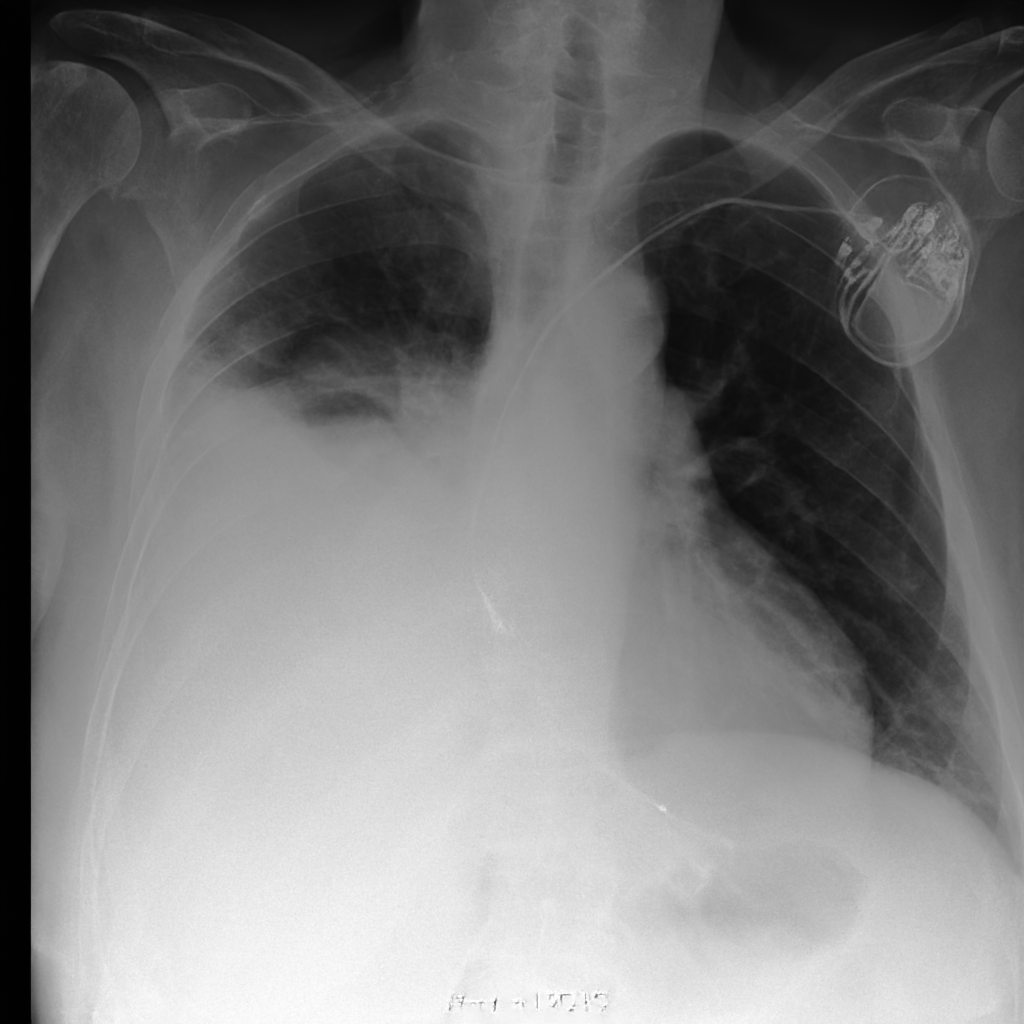} &   \includegraphics[width=0.18\textwidth]{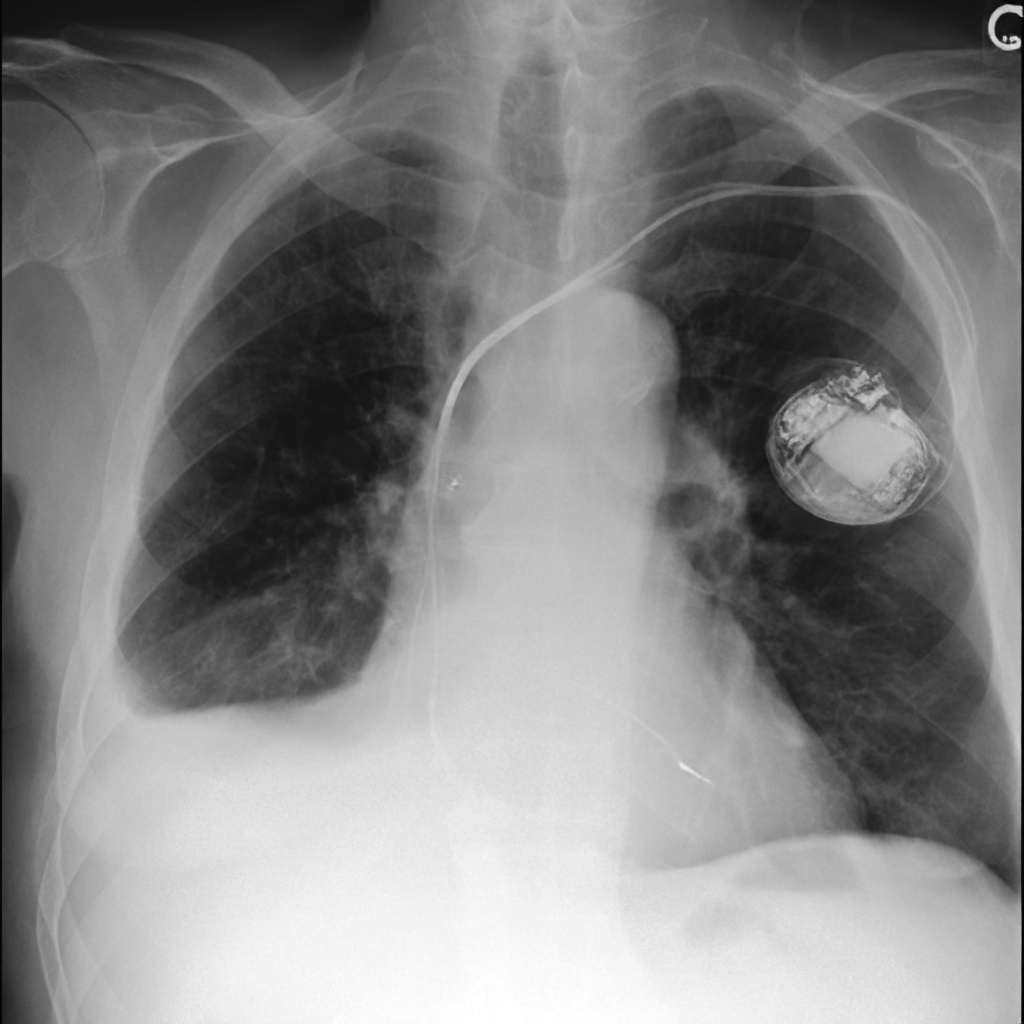} \\
    \end{tabular}
    \caption{\textit{``Large pleural effusion is in the right lower lung. A pacemaker is in the left upper chest.''}}
    \label{fig:ex-prompt-pace}
     \end{subfigure}
     \par\medskip
    \begin{subfigure}[b]{0.22\textwidth}
         \centering
  \includegraphics[width=0.75\textwidth]{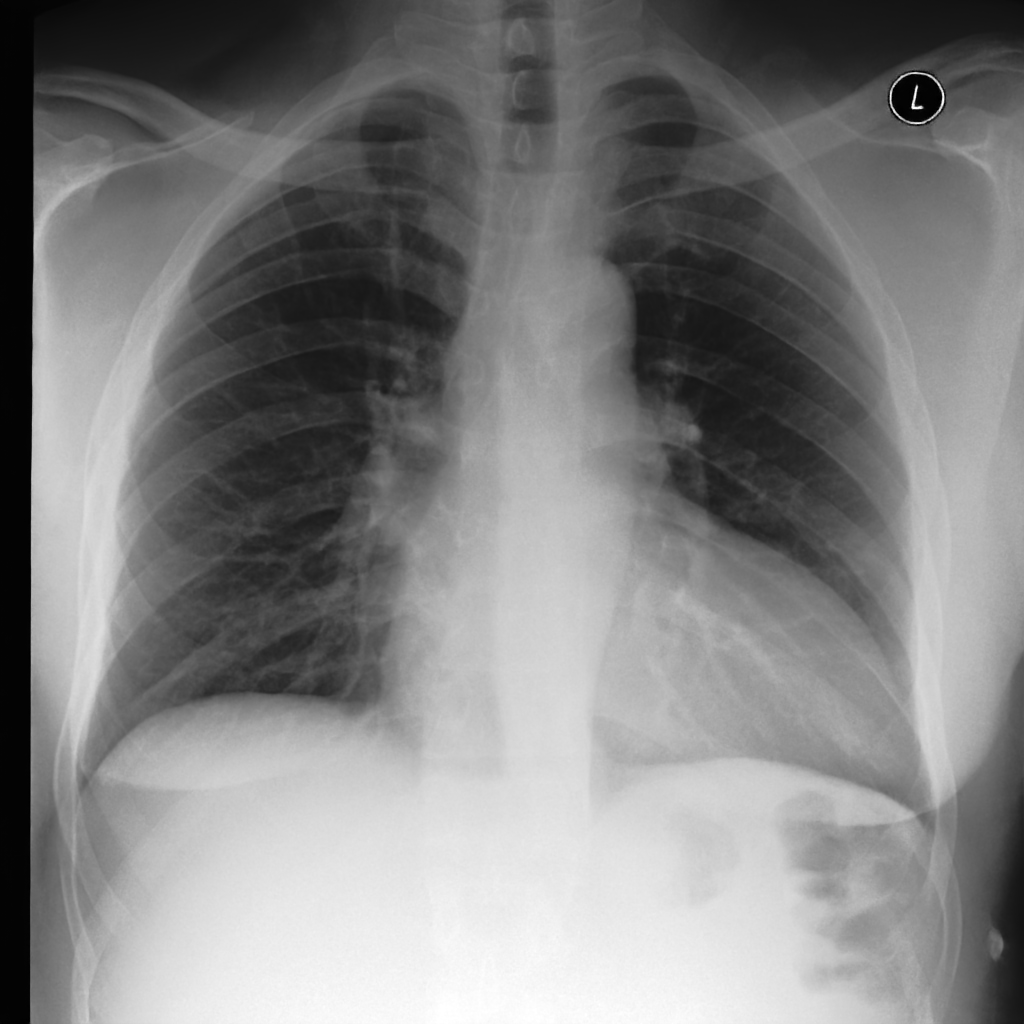}
    \caption{\textit{``Acute \\cardiomegaly.''\\ \hspace{1mm}}}
    \label{fig:ex-prompt-cardio}
     \end{subfigure}
    \hfill
    \begin{subfigure}[b]{0.22\textwidth}
         \centering
  \includegraphics[width=0.75\textwidth]{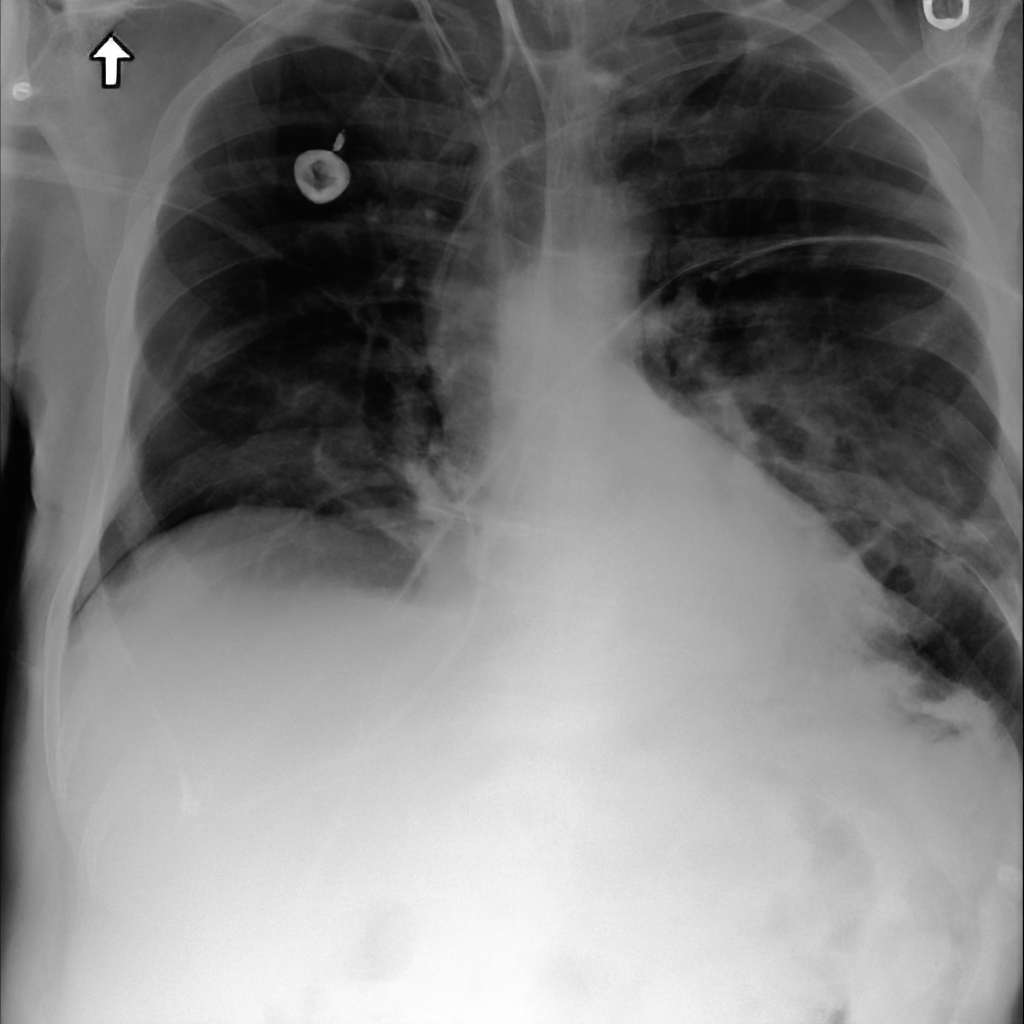}
    \caption{\textit{``Prominent \\ left-sided atelectasis.''}}
    \label{fig:ex-prompt-atelec}
     \end{subfigure}
    \hfill
    \begin{subfigure}[b]{0.22\textwidth}
         \centering
  \includegraphics[width=0.75\textwidth]{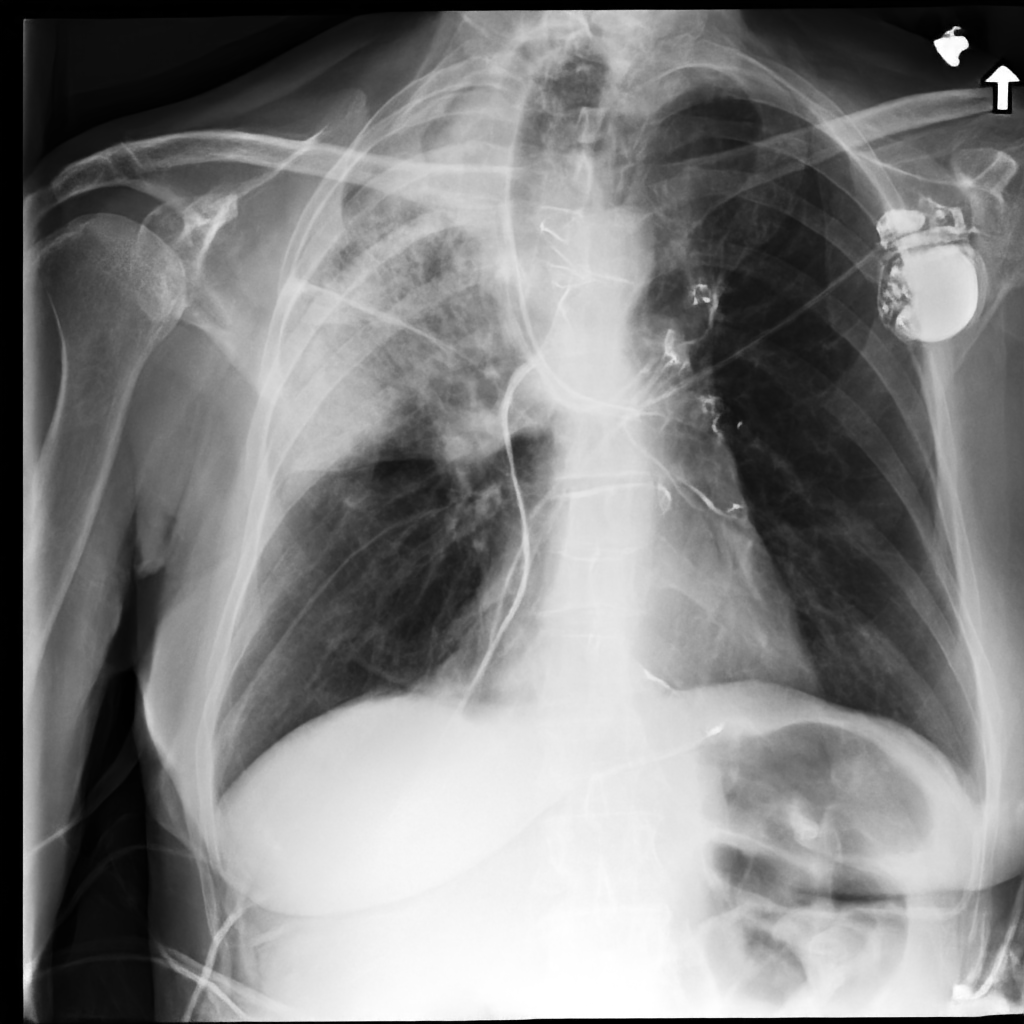}
    \caption{\textit{``Mass filling is in the upper zone of the right lung.''}}
    \label{fig:ex-prompt-mass}
     \end{subfigure}
    \hfill
    \begin{subfigure}[b]{0.22\textwidth}
         \centering
  \includegraphics[width=0.75\textwidth]{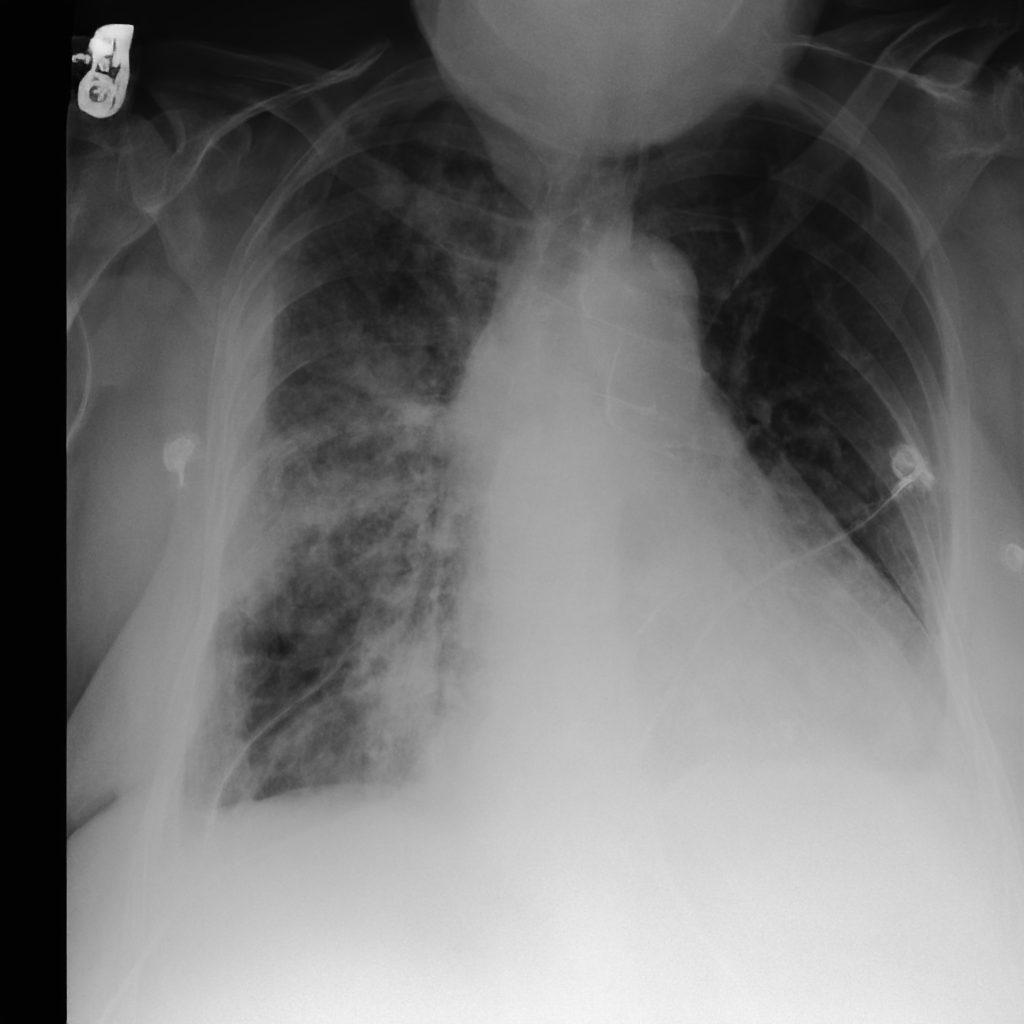}
    \caption{\textit{``Pneumonia is in the right lung.'' \\ \hspace{1mm}}}
    \label{fig:ex-prompt-pneumonia}
     \end{subfigure}
     
    \caption{Prompting a text-conditioned \textit{Cheff} via radiological findings.}
    \label{fig:ex-prompt}
\end{figure}

\paragraph{Radiological Report-to-Image Synthesis}
MIMIC-CXR provides a range of radiological text reports for every study (over 220,000), which can be utilized as conditioning $\bm y$. 
The sections \textit{Findings} and \textit{Impressions} are extracted from the raw reports and concatenated before applying a tokenizer.
Following \cite{rombach_high-resolution_2022}, $\tau_\phi$ is a trainable BERT-style encoder-only transformer.
Examples in Figure~\ref{fig:ex-prompt} show that the model has learned concepts of various pathologies and
allows for a customized synthesis of a patient's condition.
Conditioning on both \textit{Findings} and \textit{Impressions} allows controlling not only pathologies but also the creation of external devices like pacemakers.
Furthermore, the freedom of text inputs allows for specifying the localization of the targeted disease or item.
%
\section{Conclusion} \label{sec:conclusion}

In this paper, we advance the state-of-the-art in Chest X-ray synthesis by proposing a multi-stage foundational cascaded latent diffusion model called \textit{Cheff}.
A success factor for the quality of \textit{Cheff} is \textit{MaCheX} - our large-scale multi-centric Chest X-ray collection from numerous, publicly available datasets with  high diversity in phenotypes, medical conditions, diseases, and medical devices.

Approaching the terrain of indistinguishable synthetic patient samples, \textit{Cheff} requires an increased awareness of potential harms and responsible utilization.
Further, despite being able to generate realistic-looking 1-megapixel radiographs, modern X-ray scanners produce images with an up to 7-megapixel resolution and 14-bit depth, which is a requirement for radiological reading on dedicated monitors and is still out of reach for current chest X-ray synthesis.
When developing downstream applications with \textit{Cheff} for a clinical context, this should be kept in mind and potential biases should be addressed, e.g., through rigorous evaluation on real-world data.

By proposing \textit{Cheff}, we offer a high-capacity generator for chest radiographs that forms the basis for a variety of use cases, e.g., image inpainting for removal of distracting medical devices or aiding existing efforts to increase reliability in classifiers by generating underrepresented classes.
Our method does not only offer traditional synthesis but also enters the exciting area of radiological report-to-chest-X-ray generation, which allows fine-grained control over the diffusion process via text prompts.


\section*{Acknowledgments}

The authors gratefully acknowledge LMU Klinikum for providing computing resources on their Clinical Open Research Engine (CORE).
This work has been partially funded by the Deutsche Forschungsgemeinschaft (DFG, German Research Foundation) as part of BERD@NFDI - grant number 460037581.

%
%
\bibliographystyle{splncs04.bst}
\bibliography{chexray-diffusion}

\clearpage
\section{Supplementary Material}

\subsection{Training and sampling with diffusion models.}

The following two algorithms show the training and sampling strategy for a diffusion model adapted from \cite{ho_denoising_2020}.

\begin{algorithm}[H]
\SetKwInput{kwReq}{Required}

\caption{Training of a vanilla diffusion model}\label{alg:train}
\kwReq{Denoising network $\bm \epsilon_\theta$ with parameters $\theta$, forward variances $\beta_i$ with $i \in \{1, ..., T\}$}

\Repeat{converged}{
\tcp{Sample uncorrupted image from data distribution}
$\x_0 \sim q(\x_o)$

\tcp{Sample uniformly distributed timestep}
$t \sim U({1, ..., T})$

\tcp{Sample normal distributed noise}
$\bm \epsilon \sim \mathcal{N}(\bm 0, \bm I)$

\tcp{Compute alpha products}
$\bar{\alpha}_t = \prod_{s=1}^t (1-\beta_s)$

\tcp{Obtain corrupted $\x_0$ at timestep $t$}
$\x_t = \sqrt{\bar{\alpha}_t}\x_0 + \sqrt{1-\bar{\alpha}_t}\bm \epsilon$

\tcp{Update weights with loss gradients}
$\theta \gets \theta - \nabla_\theta \Vert \bm \epsilon_t - \bm \epsilon_{\theta}(\x_t, t) \Vert^2_2$
}
\end{algorithm}
\begin{algorithm}[H]
\SetKwInput{kwReq}{Required}

\caption{Sampling from a vanilla diffusion model}\label{alg:samp}
\kwReq{Denoising network $\bm \epsilon_\theta$, forward variances $\beta_i$ with $i \in \{1, ..., T\}$, reverse variances $\sigma_i$ with $i \in \{1, ..., T\}$}

 \tcp{Sample start as isotropic noise}
$\x_T \sim \mathcal{N}(\bm{0}, \bm{I})$

 \For{$t \gets T$ \KwTo $1$}{

\tcp{Compute alpha products}
$\bar{\alpha}_t = \prod_{s=1}^t (1-\beta_s)$

\tcp{Do not set any noising for the last step}
\lIf{$t > 1$}
{
    $\bm \epsilon \sim \mathcal{N}(\bm 0, \bm I)$
     \textbf{else} $\epsilon = \bm 0$
}

\tcp{Compute state of \x\ on previous timestep}
$\x_{t-1} \gets \frac{1}{\sqrt{1-\beta_t}} \left( \x_t - \frac{\beta_t}{\sqrt{1 - \bar{\alpha}_t}} \bm \epsilon_\theta(\x_t, t) \right) + \sigma_t \epsilon$

}
\Return $\x_0$
\end{algorithm}
Here, $\sigma_t$ is the variance of the reverse process at timestep $t$. 
While this parameter is learnable in general, \cite{ho_denoising_2020} presents two static versions of $\sigma_t$: Setting it directly to $\beta_t$ or $\frac{1 - \bar{\alpha}_{t - 1}}{1 - \bar{\alpha}_t - } \cdot \beta_t$.
We chose the latter option in our experiments.
We refer to \cite{ho_denoising_2020} for extended derivations and proofs for the above algorithms and a deeper understanding of diffusion modeling.


\subsection{MaCheX Collection Details}

In the following, more details about the components of \textit{MaCheX} are given.

\begin{itemize}
    \item \textbf{ChestX-ray14} \cite{wang_chestx-ray8_2017} contains 108,948 frontal-view X-ray images from the NIH clinical center in Bethesda between 1992 to 2015. Labels for over 14 pathologies are sourced via NLP methods from radiological reports.
    \item \textbf{CheXpert} \cite{irvin_chexpert_2019} consists of 224,316 frontal- and lateral-view scans from the Stanford Hospital in Palo Alto. The studies were performed between 2002 and 2017. While text reports are also the basis for the provided labels, the extraction process offers the option of marking a pathology as \textit{not mentioned} or \textit{uncertain}. The so-called \textit{CheXpert labeler} is also open-sourced, but the text reports are not.
    \item \textbf{MIMIC-CXR} \cite{johnson_mimic-cxr_2019} provides 377,110 images from the Beth Israel Deaconess Medical Center Emergency Department recorded between 2011 and 2016. In addition to labels provided in the same style as in CheXpert, the dataset also releases full-text radiological reports.
    \item \textbf{PadChest} \cite{bustos_padchest_2020} contains studies with 160,868 images from the San Juan hospital in Alicante from 2009 to 2017. The dataset promotes radiological texts in Spanish and inferred labels from the reports.
    \item \textbf{BRAX} \cite{reis_eduardo_pontes_brax_nodate} is a Brazilian dataset from the Hospital Israelita Albert Einstein, housing 40,967 chest x-rays with supplied labels in the style of CheXpert.
    \item \textbf{VinDr-CXR} \cite{nguyen_vindr-cxr_2022} consists of 18,000 thoracic scans from 2018 to 2020 recorded at Hospital 108 and Hanoi Medical University Hospital in Vietnam. It provides hand-labeled annotations with coordinate bounding boxes for various pathologies.
\end{itemize}


\subsection{Training details of Cheff}

All model training was conducted on an NVIDIA DGX system with 8 A100 @ 40GB VRAM.
To reduce computational complexity, we do not conduct a hyperparameter search but rely on the recommended and subsequently elaborated settings proposed in \cite{rombach_high-resolution_2022, saharia_image_2021}.
To accelerate the sampling procedure in the diffusion models, we utilize DDIM sampling with 150 steps.
All models were trained with the Adam optimizer.

\subsubsection{Autoencoder (AE)}

For AE we use the KL-regularized continuous variational AE (VAE) with a downsampling factor 4 as in \cite{rombach_high-resolution_2022} and train it 3 epochs with a batch size of 64 on \textit{MaCheX}.
We did not see an improvement in validation errors in longer training sessions.
The convolutional embedding has a dimension of $3 \times 64 \times 64 = 12288$.
The architecture consists of 3 blocks with 128, 256, and 512 channels respectively.
Each block contains two residual layers.
No self-attention or dropout was applied.
The decoder is the mirrored version of the encoder.
To limit the normal prior in the latent space and focus on reconstruction quality, a KL-divergence weighting of \num{1e-6} was set.
We stick to the default learning rate of \num{4.5e-6}.
We experimented with larger downsampling factors as well as discretized embeddings and found that factor 4 offers the highest, most fine-grained reconstruction quality.

\subsubsection{Semantic diffusion model (SDM)}

\paragraph{Unconditional SDM}
The model is trained on the full \textit{MaCheX} collection for 500,000 steps with a batch size of 256.
The time-conditional U-net has 4 downsampling blocks with a base filter number of 224 and the multipliers $(1, 2, 4, 4)$ respectively.
Each downsampling block has 2 residual layers.
Attention is applied to the spatial resolution of 8, 16, and 32.
The learning rate is set to \num{5.0e-05}.
The timestep is encoded over sinusoidal position embeddings.

We noticed that the default linear variance schedule with a maximum variance of 0.0195 does not induce enough noise into the process to guarantee an approximate isotropic  Gaussian in $\x_T$, which proves to be extremely destructive for the synthesis process.
The number of used timesteps is 1000.
Raising the maximum variance to 0.0295 eliminates this issue.

\paragraph{Report-conditioned SDM}
The structure and training procedure of the base U-net stays the same, but we train for a longer period of time, namely for 750,000 steps.
We apply the BERT-tokenizer to preprocess the text sequences and limit the length of the tokenized input to a maximum of 150. 
The conditioning $\bm y$ is fed to the model over cross-attention, where $\tau_\phi$ is a transformer with a depth of 32 and an embedding dimension of 1280.
As it is described in \cite{rombach_high-resolution_2022, vaswani_attention_2017}, an embedding $\tau_\phi(\bm y) \in \mathbb{R}^{d_{\tau}}$ is induced to the model over cross-attention.
Thus, an attention layer with $\text{Attention}(Q,K,V) = \text{softmax} \left( \frac{QK^T}{\sqrt{d}} \right) V$ has keys $K$, queries $Q$ and values $V$ defined as
\begin{align*}
Q &= W_Q \cdot \varepsilon  &  K &= W_K \cdot \tau_\phi(\bm y)  &  V &= W_V \cdot \tau_\phi(\bm y)    \eqendp\\
\end{align*}
$W_Q \in \mathbb{R}^{d_q \times d_\varepsilon}, W_K \in \mathbb{R}^{d_k \times d_\tau}, W_V \in \mathbb{R}^{d_v \times d_\tau}$. $\varepsilon$ is an intermediate vectorized representation in $\bm \epsilon_\theta$, i.e., the input to the respective attention layer in the time-conditioned denoising U-net.
In other words, cross-attention works by providing keys $K$ and values $V$ with information over the targeted conditioning and querying (via $Q$) with the concept that is being processed, i.e., the image to be denoised. 

\subsubsection{Super-resolution (SR)}
SR training amounts to 1 million steps on \textit{MaCheX} with a batch size of 32.
The conditioning is done by simply concatenating the low-resolution image to the target input on the channel dimension.
To obtain \xlr\ for training, we resize \xhr\ to the low resolution with bicubic downsampling.
To align SR with the synthesis pipeline, we fine-tune SR for another 500,000 steps on $D(\z)$.
The learning rate for normal training was \num{5.0e-05} and \num{2.0e-05} during finetuning.
A cosine variance schedule with 2000 timesteps is used instead of the piece-wise distribution in \cite{saharia_image_2021}.

The model itself is a U-net with sinusoidal position embeddings for the timestep.
It has a base number of filters of 16 and 8 downsampling blocks following the filter multiplication schedule of $(1, 2, 4, 8, 16, 32, 32, 32)$.
Each block has 2 residual layers.
Self-attention is applied in the lowest 5 downsampling levels.


\subsection{Sparse latent space}

While the basic formulation of a VAE often promises the existence of a generative model in itself, we observe that this is not the case in the AE component here.
As the AE is trained with an extremely low weighting of KL-divergence, the influence of the Gaussian prior is minimal in the structure of the posterior latent space $\mathcal{Z}$.
As a result, $\mathcal{Z}$ is not smooth, and sampling $\z \sim \mathcal{N}(\bm{0}, \sigma \bm{I})$ does not yield any valid samples of the dataspace even for small values of $\sigma$.
This can also be easily explained, as the convolutional AE fosters a spatial embedding, and isotropic sampled noise does not resemble anything lung shape-related.
Yet, simple linear interpolation between two valid samples in latent space still seems rather smooth, but when investigated in detail, the transition is more a blending of the two images instead of merging on a conceptual level.
\begin{figure}
    \centering
    \includegraphics[width=\textwidth]{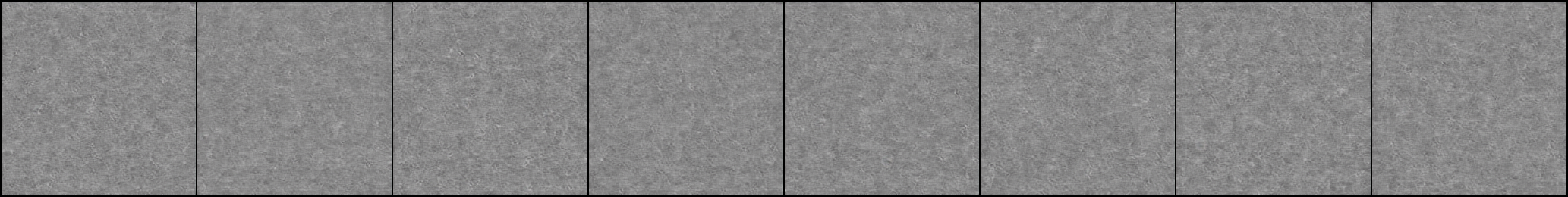}
    \caption{Reconstructions for samples $\z \sim \mathcal{N}(\bm{0}, \bm{I})$ showing missing generative capacity in a non-smooth latent space for the AE without a functioning prior like a diffusion model.}
    \label{fig:noise}
\end{figure}
\begin{figure}
     \centering
    \begin{tabular}{c}
  \includegraphics[width=\textwidth]{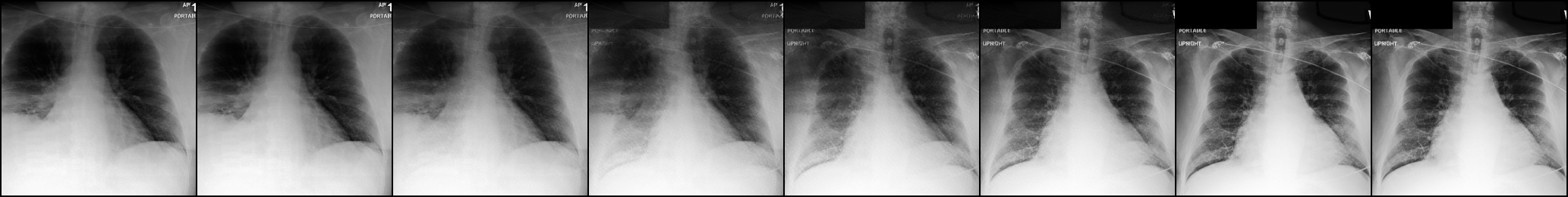} \\
  \includegraphics[width=\textwidth]{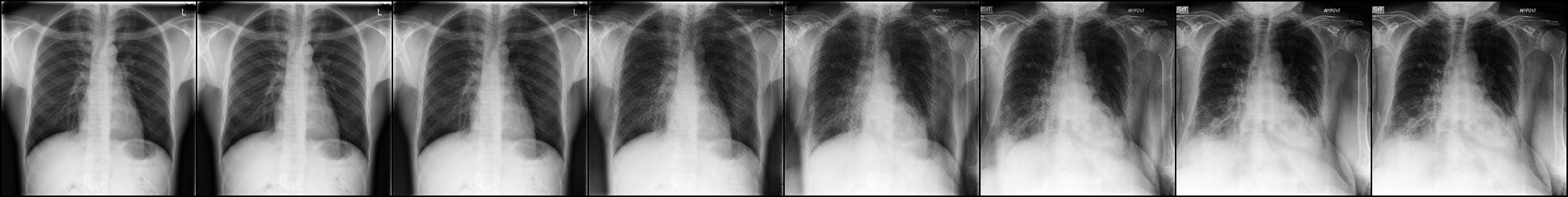} \\
  \includegraphics[width=\textwidth]{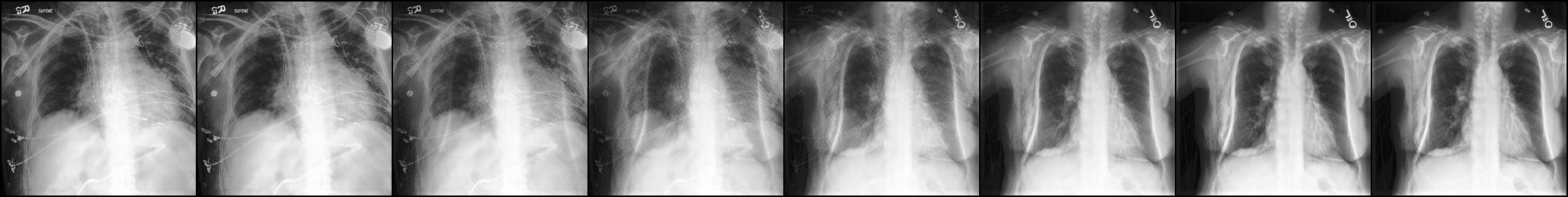} \\
    \end{tabular}
\caption{Linear interpolation in the latent space of the AE.}
\label{fig:interpol}
\end{figure}

\clearpage

\subsection{Experiments with VQ-VAE}

We also experimented with a vector-quantized version of AE, but found a dip in reconstruction quality and especially in annotations, and favored the classic VAE formulation.

\begin{figure}
     \centering
     \begin{subfigure}[b]{0.48\textwidth}
         \centering
         \includegraphics[width=\textwidth]{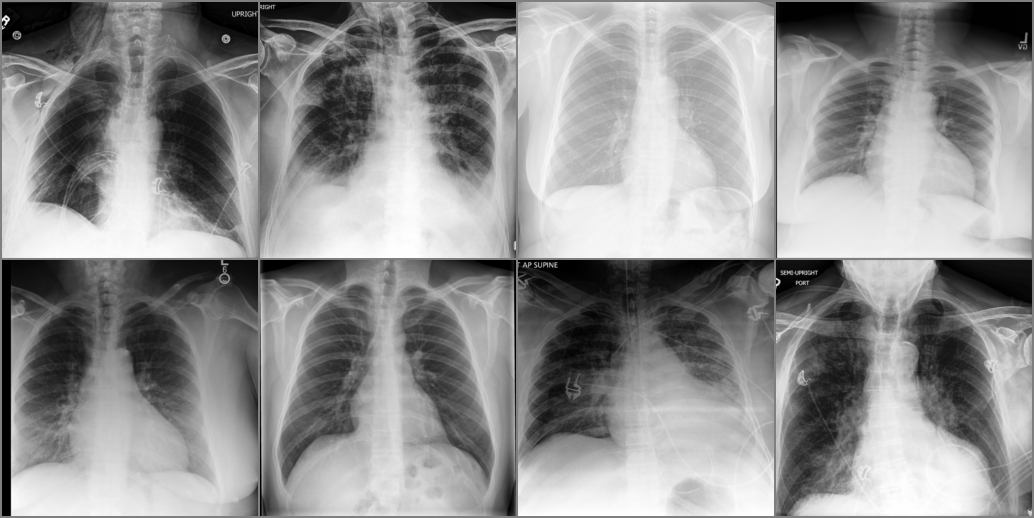}
         \caption{Original images}
     \end{subfigure}
     \hfill
     \begin{subfigure}[b]{0.48\textwidth}
         \centering
         \includegraphics[width=\textwidth]{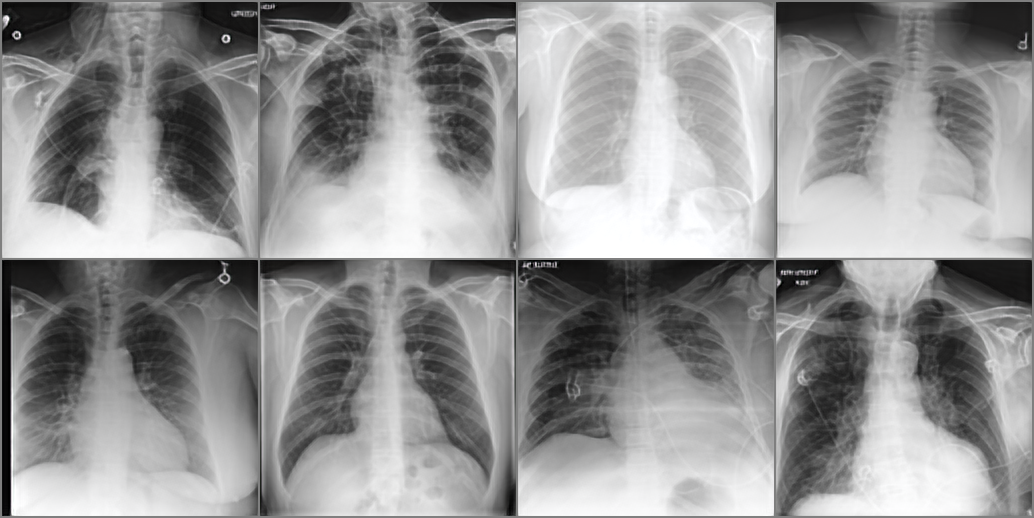}
         \caption{Reconstructed images}
     \end{subfigure}
     \hfill
        \caption{Comparison of original and reconstructed chest x-ray samples using VQ-VAE.}
\end{figure}


\subsection{Effects of insufficient noising}

\begin{figure}[b]
    \centering
    \includegraphics[width=0.9\textwidth]{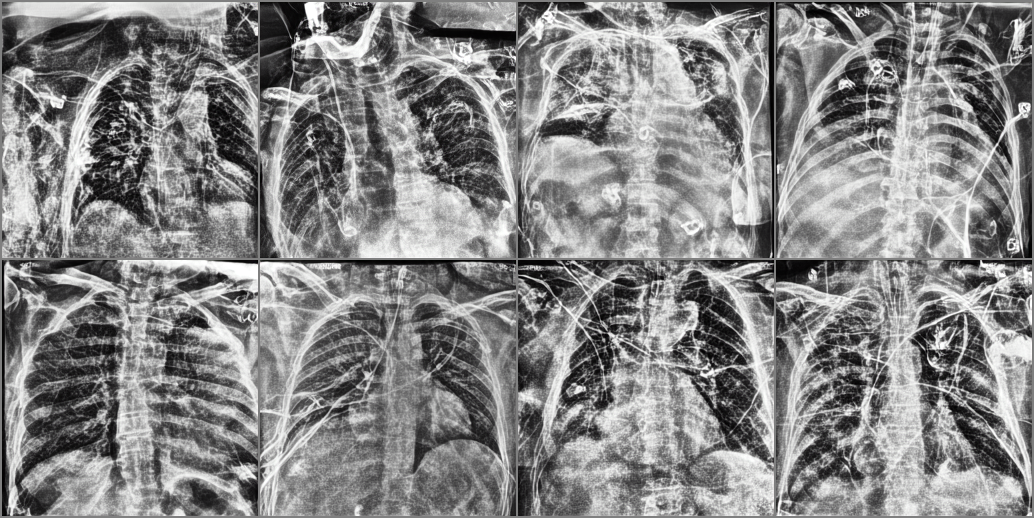}
    \caption{Insufficient noise in the forward process leads to corrupted synthesis.}
    \label{fig:noise_rec}
\end{figure}

\begin{figure}
     \begin{subfigure}[b]{0.48\textwidth}
         \centering
         \includegraphics[width=\textwidth]{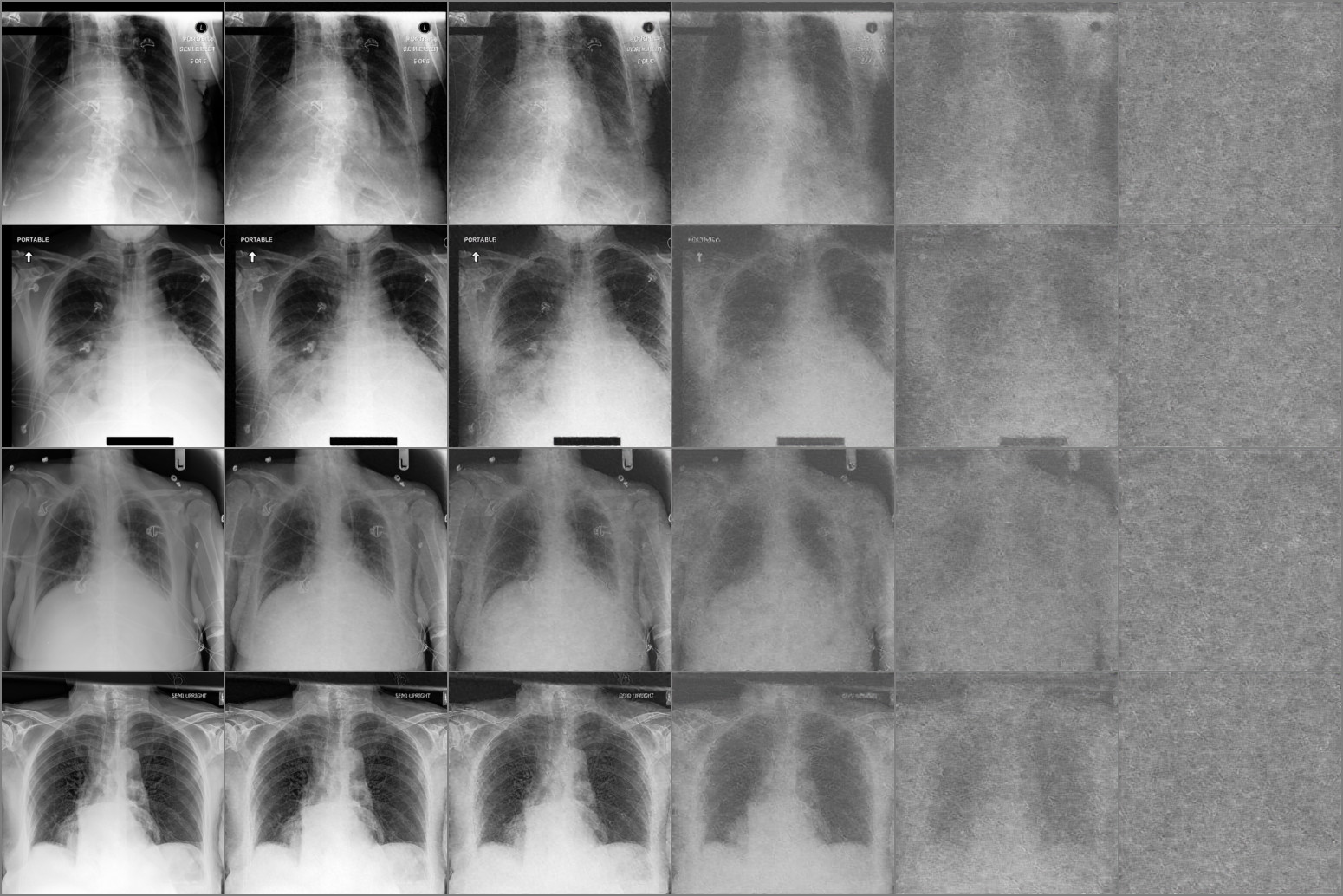}
         \caption{Low maximal $\beta$}
         \label{fig:low-beta}
     \end{subfigure}
     \hfill
     \begin{subfigure}[b]{0.48\textwidth}
         \centering
         \includegraphics[width=\textwidth]{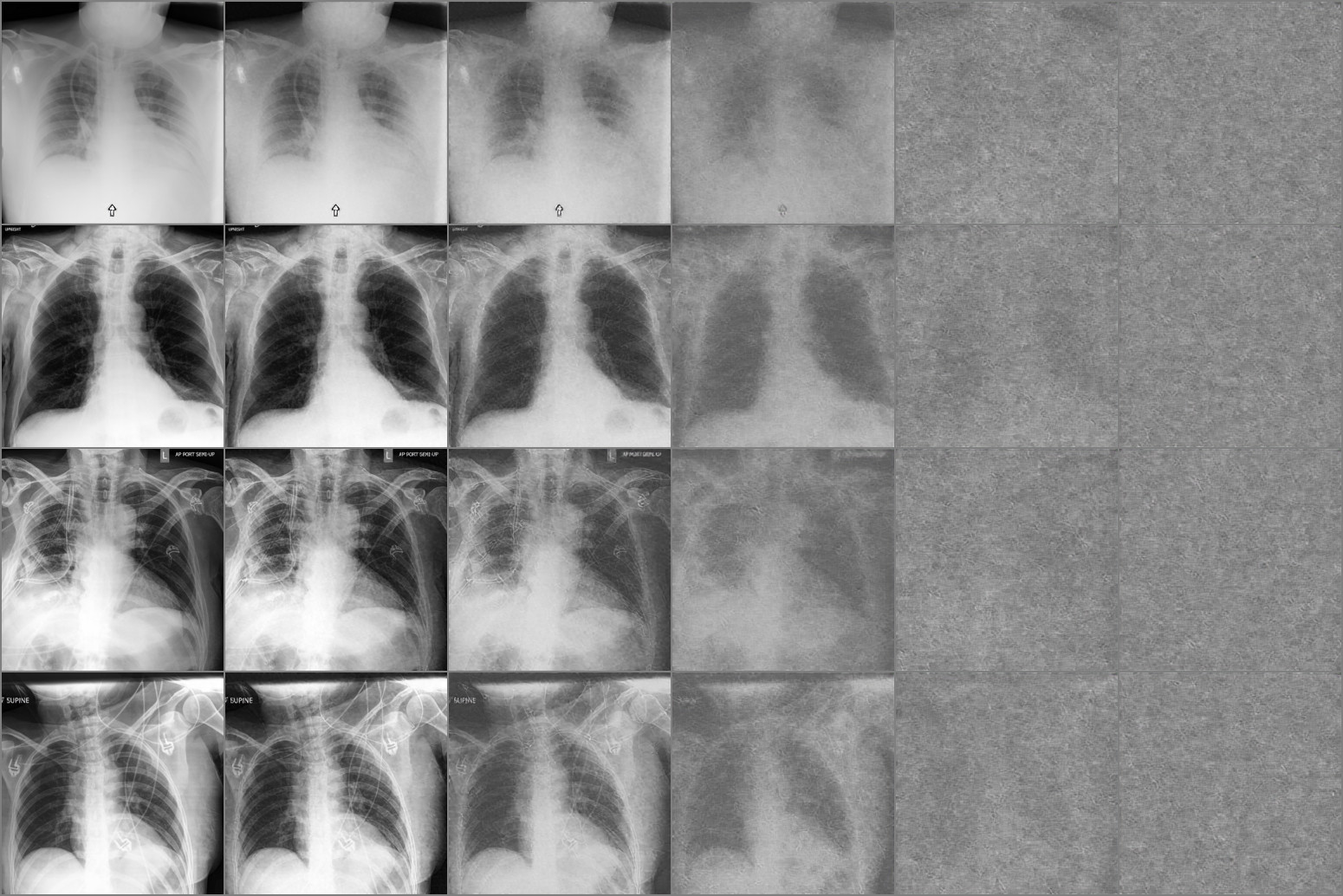}
         \caption{Increased maximal $\beta$}
          \label{fig:high-beta}
     \end{subfigure}
     \hfill
        \caption{Comparison the forward diffusion process for the default $\beta$ (\textbf{right}) and our increased value (\textbf{left}).}
\end{figure}

When training with the full default settings of LDM, we were only able to produce corrupted and solarized synthetic samples (Figure~\ref{fig:noise_rec}).
As it turns out, this was due to a too-small maximal $\beta$ value of $0.0195$ in the linear variance schedule.
By investigating the forward diffusion process, we observed that the maximal corrupted latent space sample $\z_T$ still encodes some structural information of its original image $\x_0$ (see Figure~\ref{fig:low-beta} right column) when decoding $\z_T$ with $D$.
During synthesis, where $\z_T \sim \mathcal{N}(\bm{0}, \bm{I})$ is chosen as a starting point, the model does not know what to infer from this unseen unconditional information, which results in low-quality samples.
Increasing the maximal $\beta$ to $0.0295$ resolves this issue.
%


\subsection{More inpainting examples using Cheff}

The below figure displays two more examples of inpainting in chest radiographs. The first image shows the removal of a chest port, the second image shows the successful completion of separated edges.
\begin{figure}
    \centering
    \begin{tabular}{cccc}
     \includegraphics[width=0.24\textwidth]{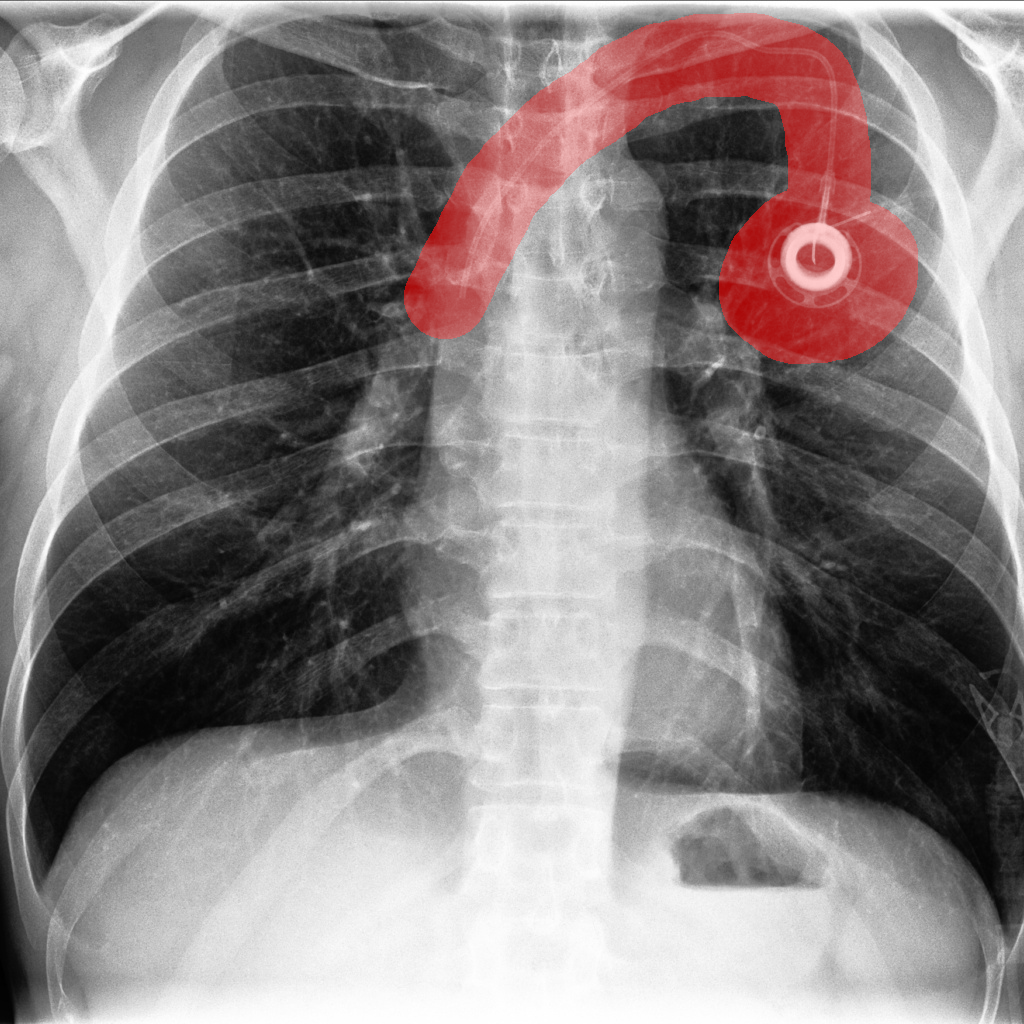} &   \includegraphics[width=0.24\textwidth]{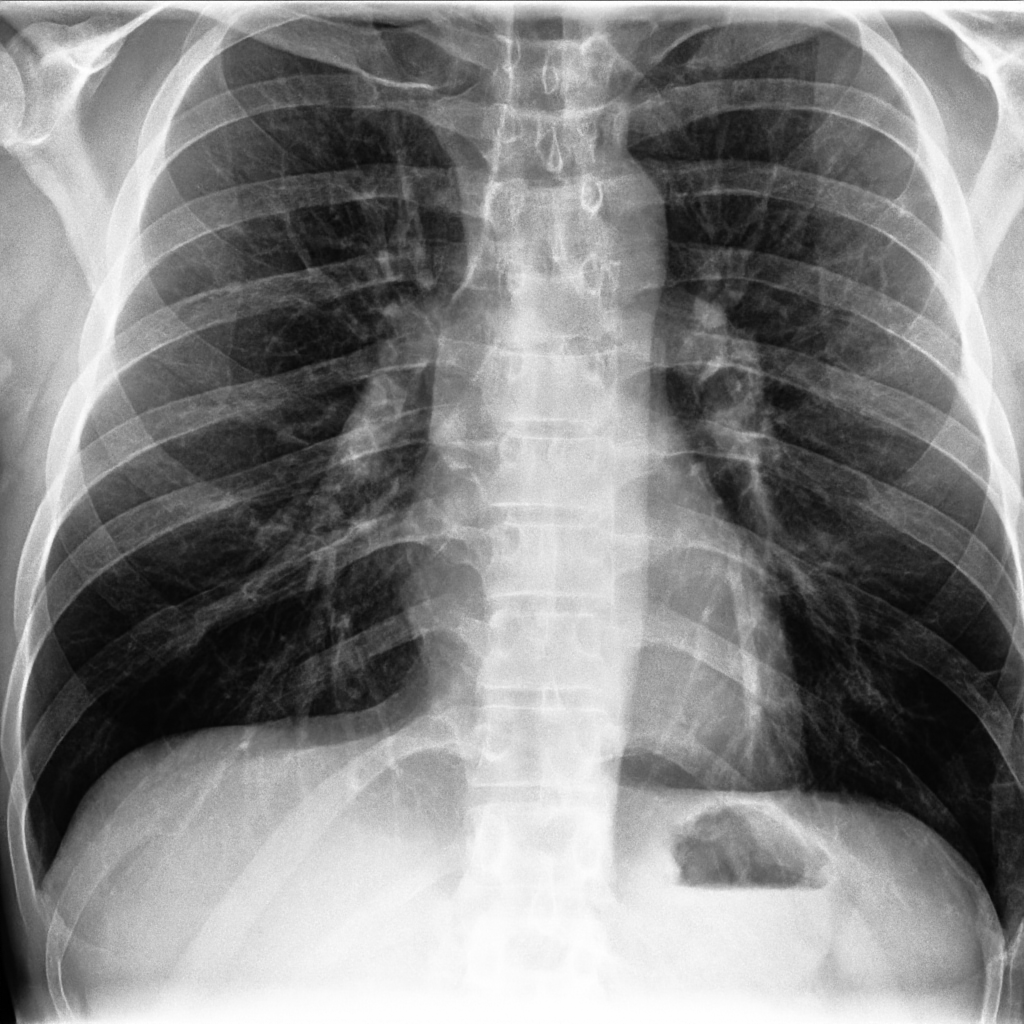} &
  \includegraphics[width=0.24\textwidth]{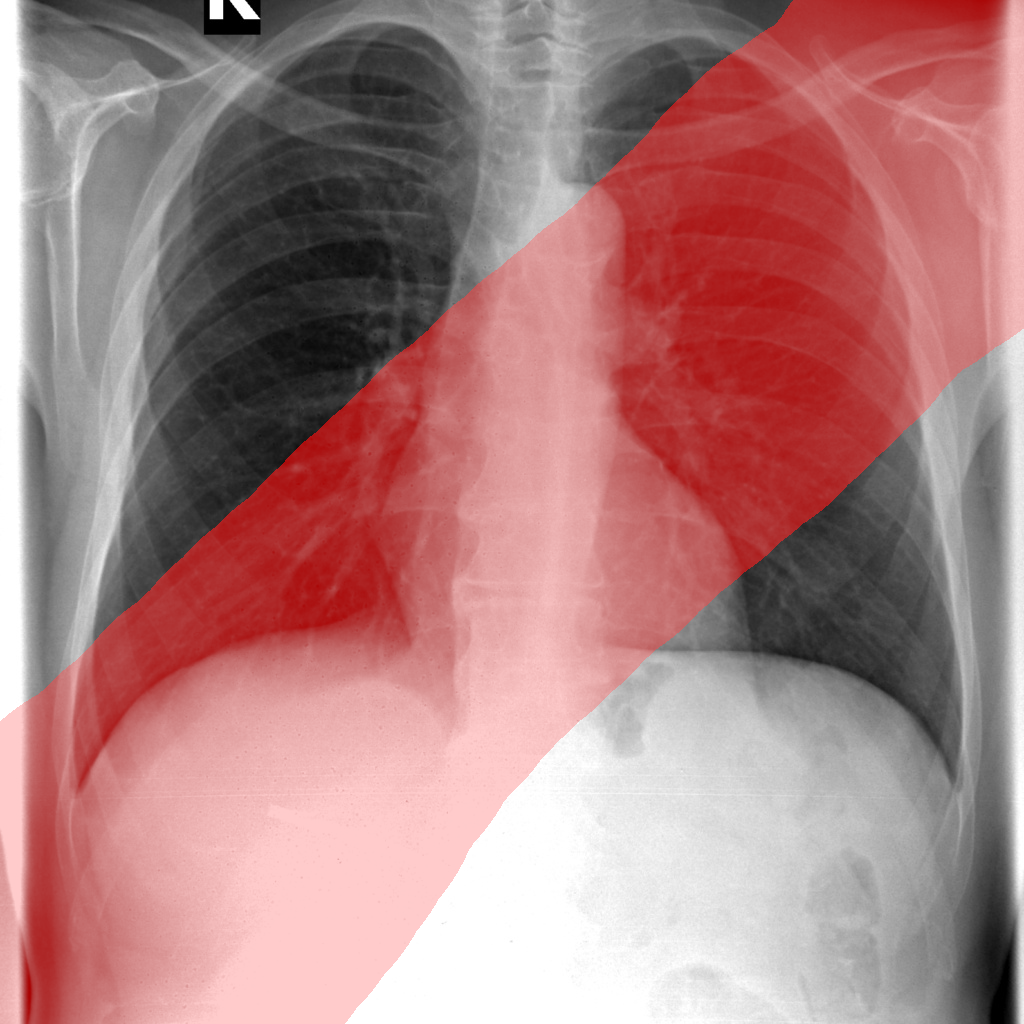} &   \includegraphics[width=0.24\textwidth]{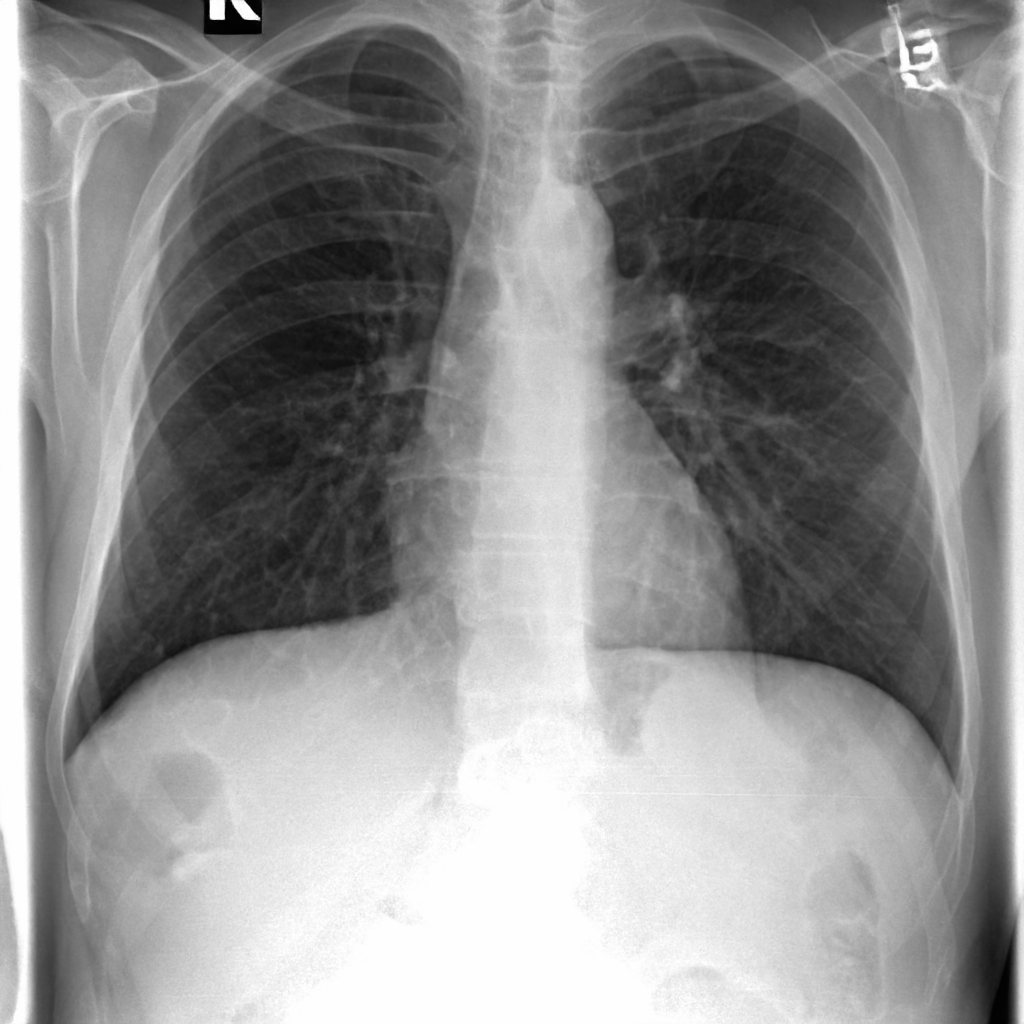} \\
    \end{tabular}
    \caption{More examples of image inpainting with \textit{Cheff}.}
    \label{fig:ex-inpaint2}
\end{figure}
%


\subsection{More synthetic samples of Cheff}

On the following pages more random synthetic Chest X-rays, which are generated by \textit{Cheff}, are showcased.

\newcommand{\RNum}[1]{\uppercase\expandafter{\romannumeral #1\relax}}

\begin{figure}
    \centering
    \begin{tabular}{c}
    \includegraphics[width=0.75\textwidth]{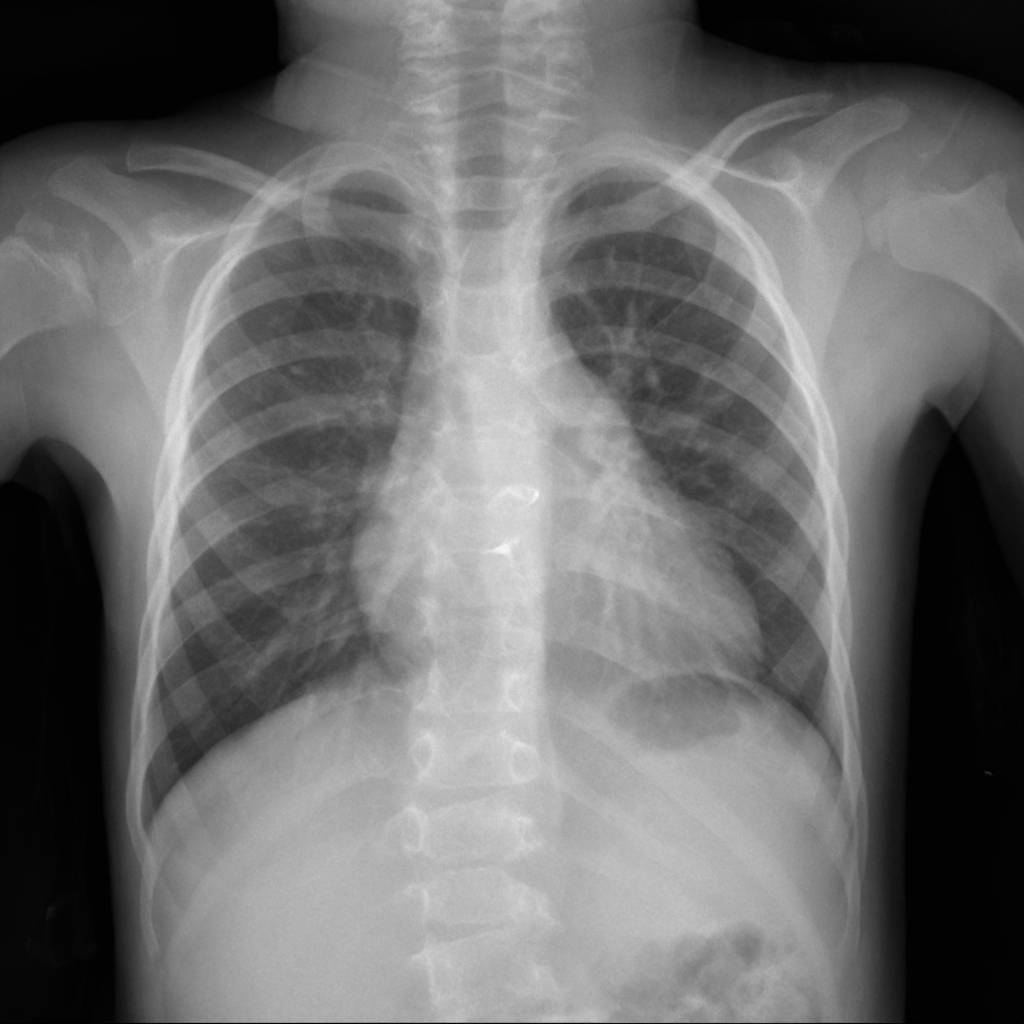} \\
    \includegraphics[width=0.75\textwidth]{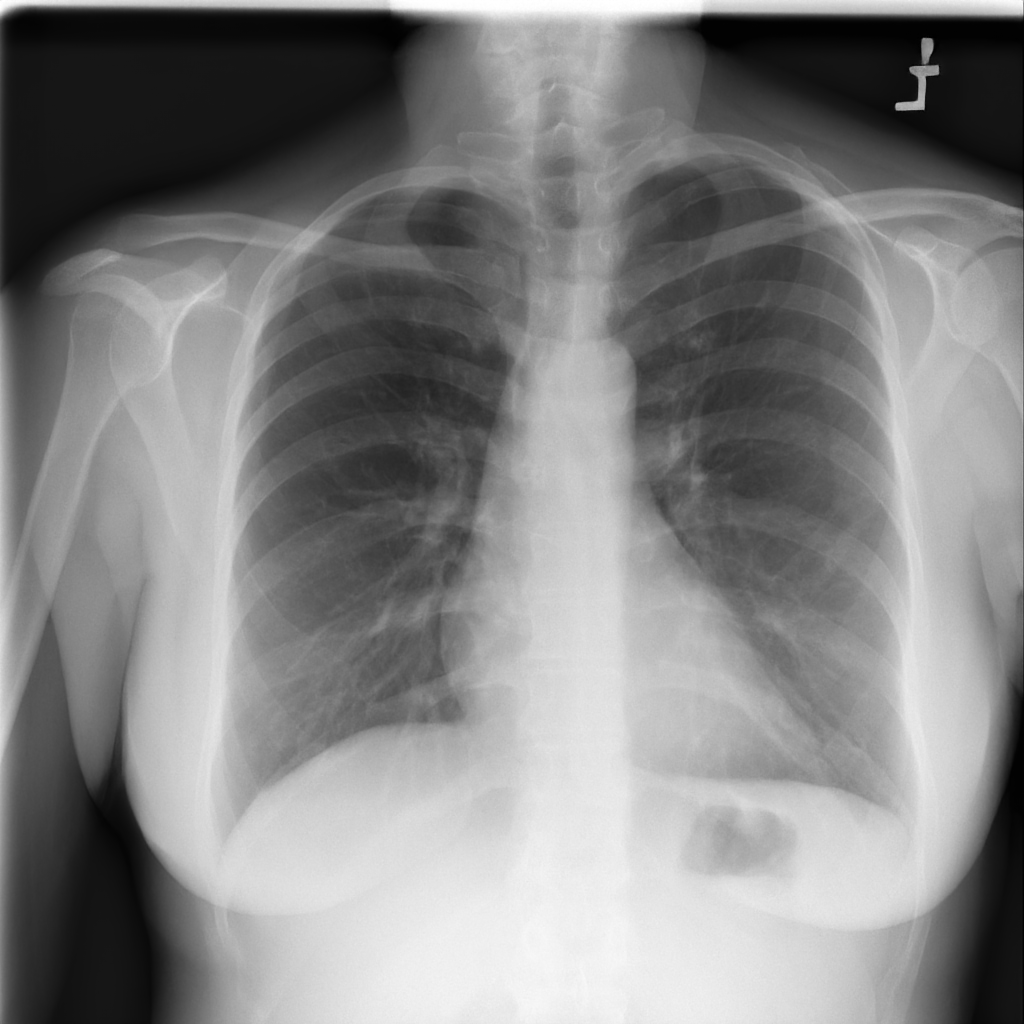} \\
    \end{tabular}
    \caption{Synthetic Chest X-rays generated by \textit{Cheff} (\textbf{\RNum{1}})}
\end{figure}
\begin{figure}
    \centering
    \begin{tabular}{c}
    \includegraphics[width=0.75\textwidth]{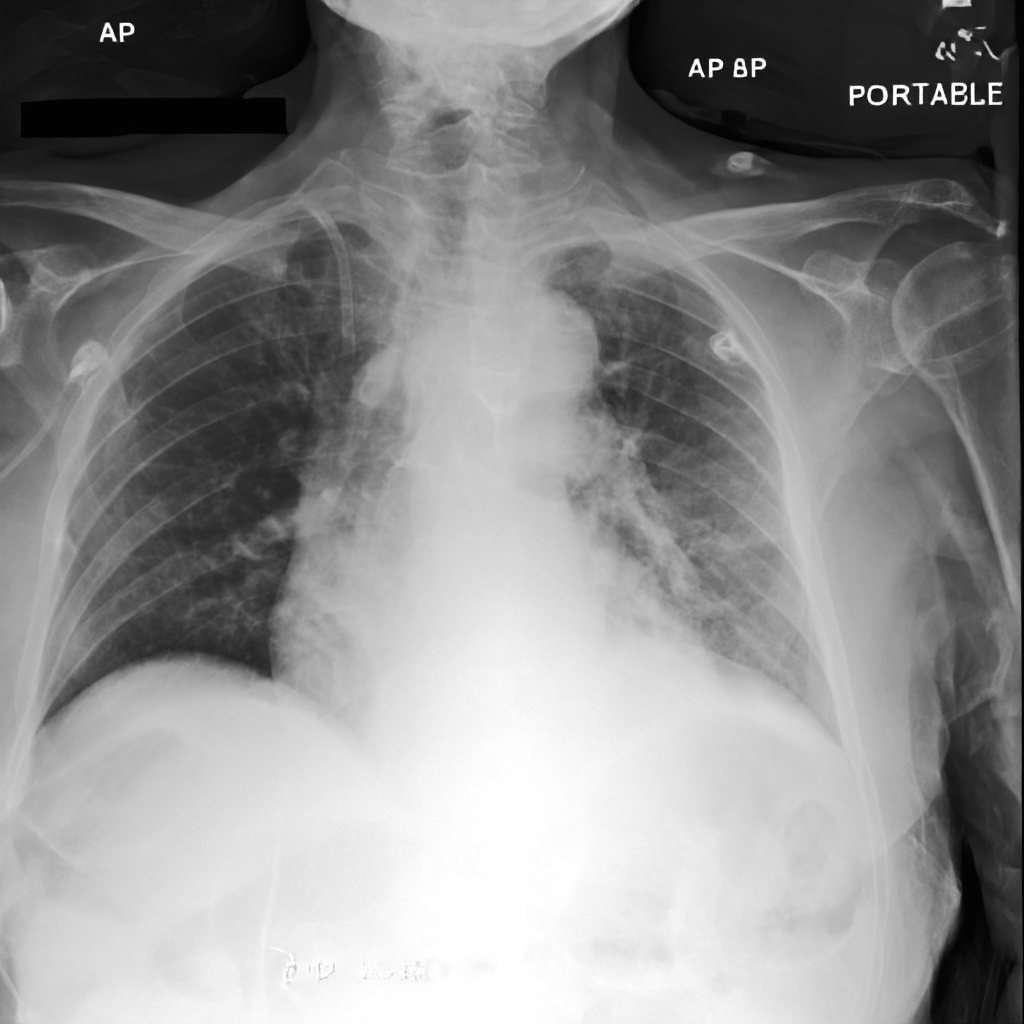} \\
    \includegraphics[width=0.75\textwidth]{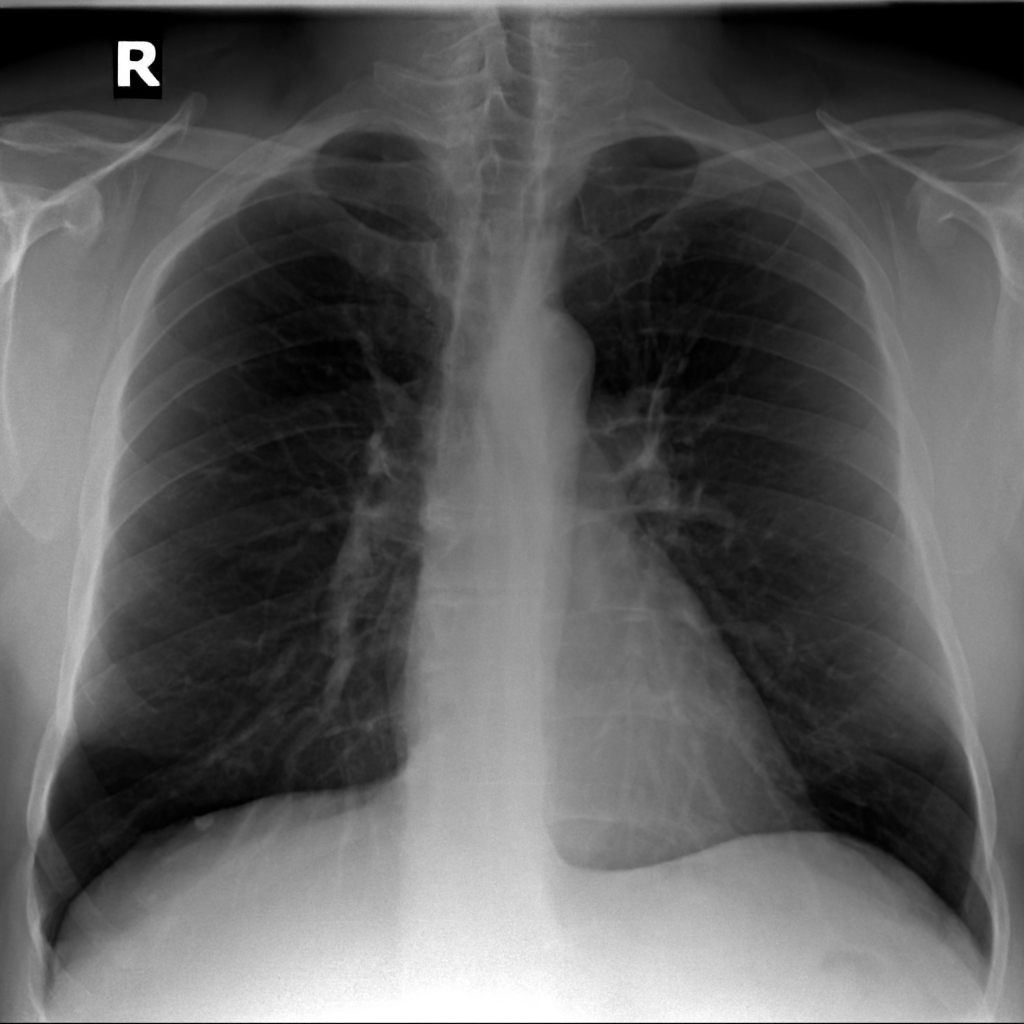} \\
    \end{tabular}
    \caption{Synthetic Chest X-rays generated by \textit{Cheff} (\textbf{\RNum{1}})}
\end{figure}
\begin{figure}
    \centering
    \includegraphics[width=0.75\textwidth]{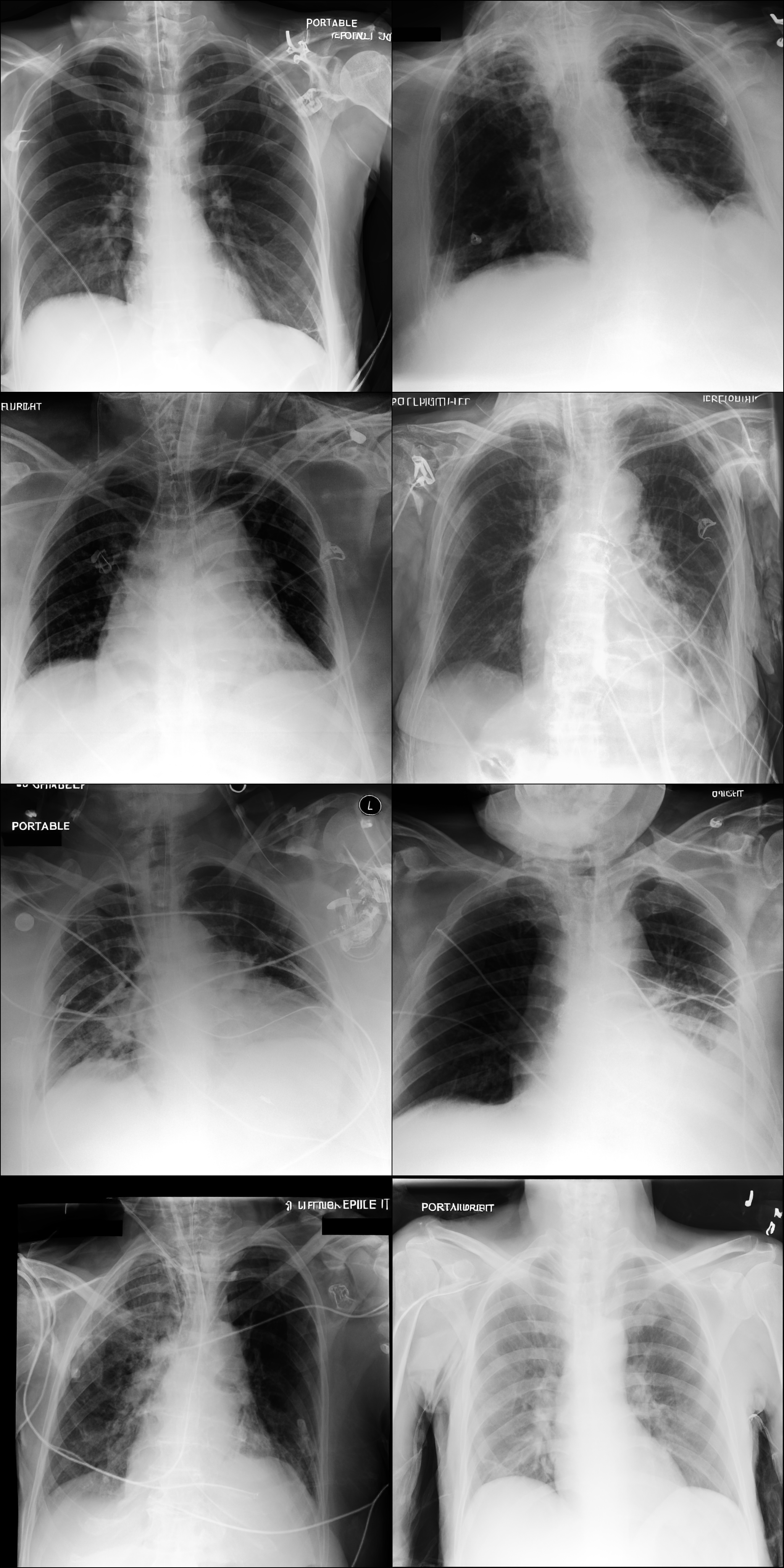}
    \caption{Synthetic Chest X-rays generated by \textit{Cheff} (\textbf{\RNum{3}})}
    \label{fig:ex-samples-l}
\end{figure}
\begin{figure}
    \centering
    \includegraphics[width=0.75\textwidth]{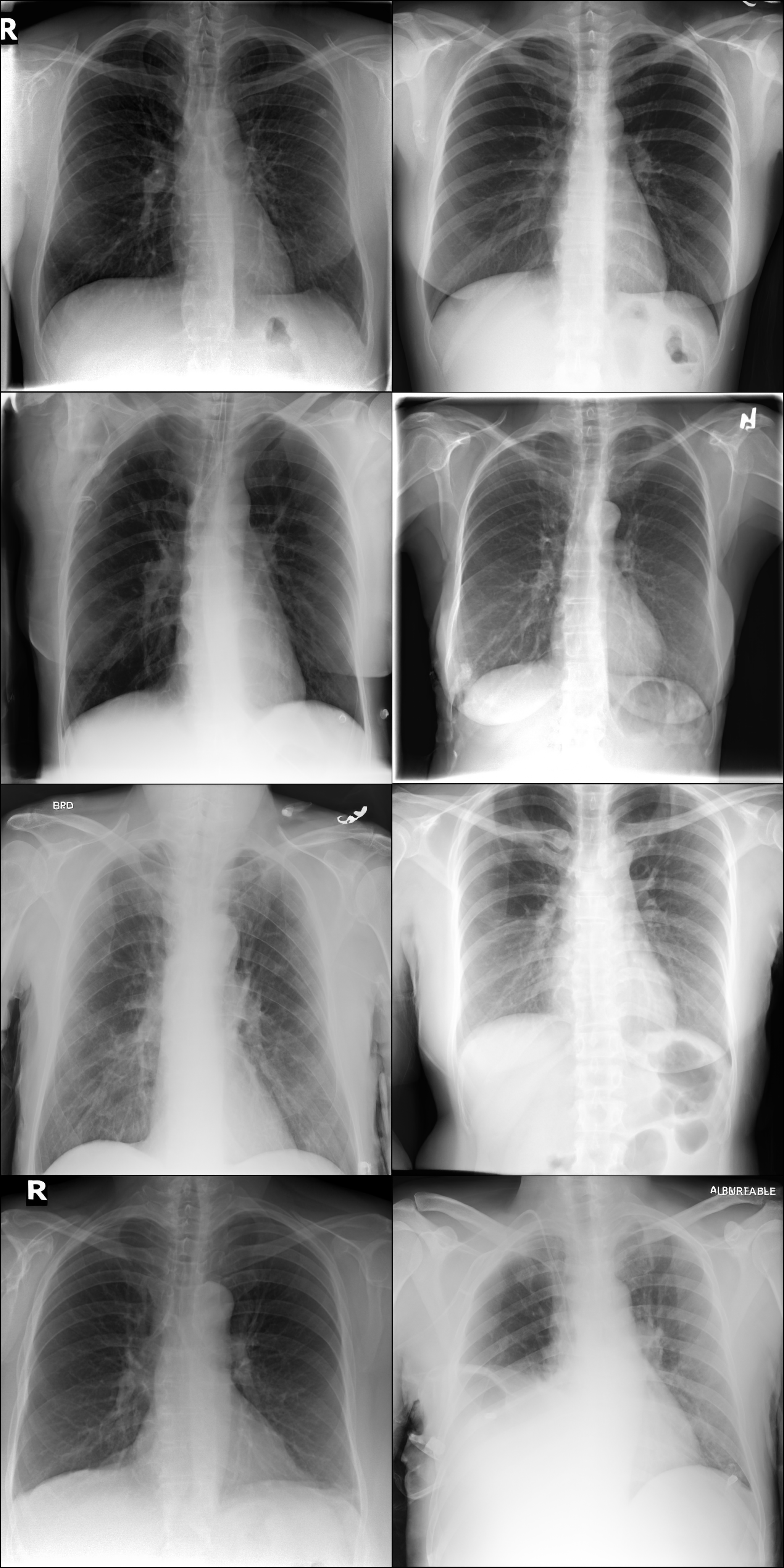}
    \caption{Synthetic Chest X-rays generated by \textit{Cheff} (\textbf{\RNum{4}})}
\end{figure}
\begin{figure}
    \centering
    \includegraphics[width=0.75\textwidth]{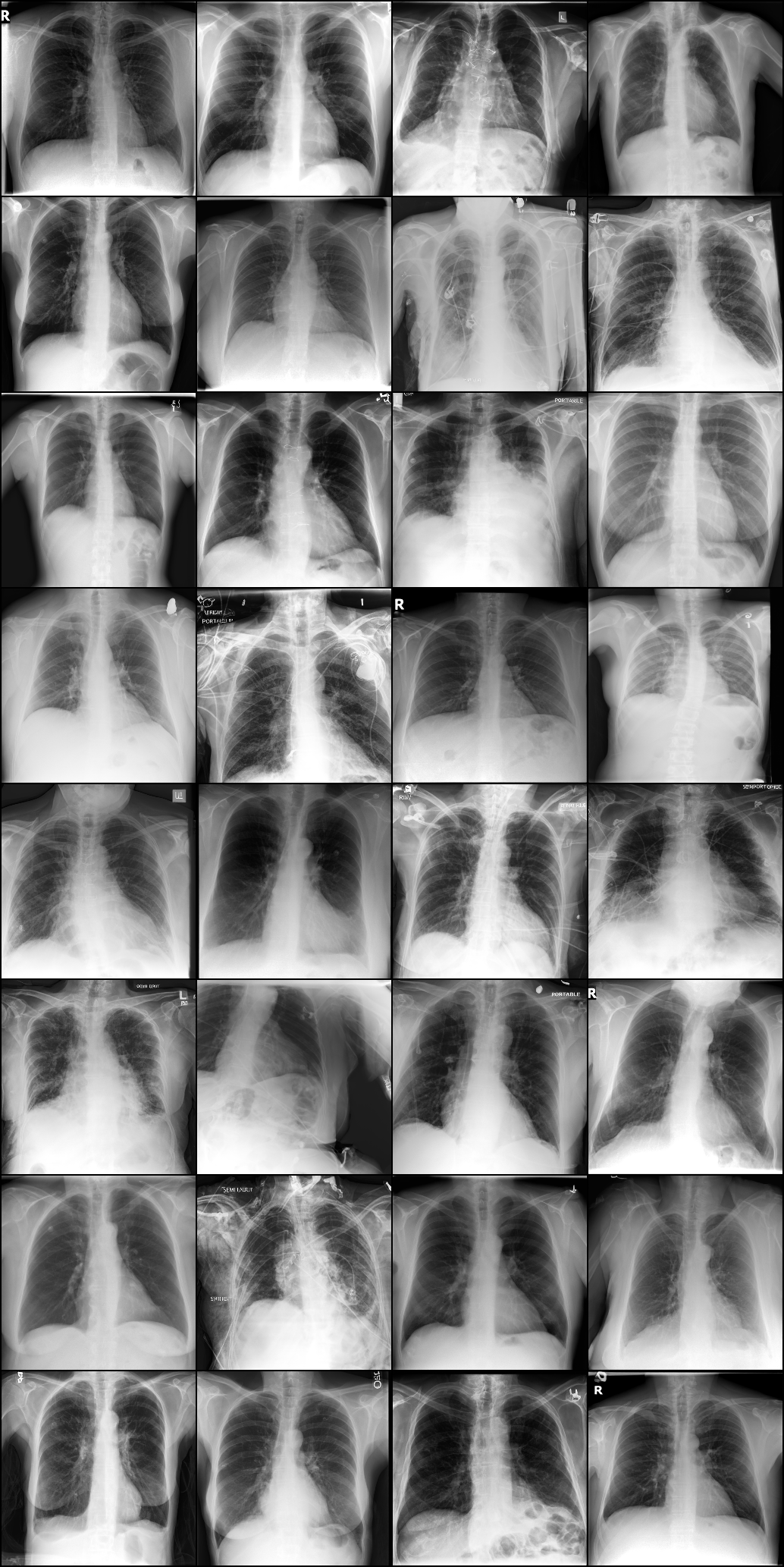}
    \caption{Synthetic Chest X-rays generated by \textit{Cheff} (\textbf{\RNum{5}})}
\end{figure}

\end{document}